\documentclass[letterpaper, preprint, paper,11pt]{AAS}	

\usepackage{bm}
\usepackage{amsmath}
\usepackage{comment}
\usepackage{textcomp}

\usepackage[colorlinks=true, pdfstartview=FitV, linkcolor=black, citecolor= black, urlcolor= black]{hyperref}
\usepackage{overcite}
\usepackage{footnpag}			      	

\usepackage{graphicx}
\usepackage{subcaption}
\usepackage{epstopdf}
\AppendGraphicsExtensions{.tif}
\usepackage{algorithm}
\usepackage{algpseudocode}
\usepackage{pgfgantt}
\usepackage{svg}
\usepackage{multirow}
\usepackage{hyperref}

\PaperNumber{23-403}

\begin{document}
\raggedbottom

\title{Estimation of inertial properties of a rigid structure maneuvered by satellite modules}

\author{Deep Parikh \thanks{Graduate Research Assistant,  Land, Air and Space Robotics (LASR) Laboratory, Aerospace Engineering}, 
Manoranjan Majji\thanks{Assistant Professor, Director, Land, Air and Space Robotics (LASR) Laboratory, Aerospace Engineering} \\
\textit{Texas A \& M University, College Station, TX, 77843-3141}}

\maketitle{}

\begin{abstract}
The LASR Laboratory is investigating the use of free-flying spacecraft modules in several on-orbit, servicing and manufacturing (OSAM) activities. Previous work consists of the system development and testing of the aforementioned thrust-capable modules. This study makes advancements to devise, implement and validate an algorithm for the estimation of inertial parameters of a rigid structure, to be maneuvered with the help of Transforming Proximity Operations and Docking Service (TPODS) satellite modules. The primary contribution of this activity is observability analysis to infer a conducive input sequence for estimating the inertial parameters. For the experimental validation of proposed estimation algorithm, real-time pose measurements are logged through the VICON® motion capture system and the recorded data is utilized to assess the performance of the estimation algorithm to predict mass and moment of inertia of an isolated TPODS module.

\end{abstract}

\section{Introduction}
Some of the initial work such as \cite{525802} focuses on the multi-robot task allocation strategies for rigid structures. Although, the formulation presented in this article can handle a vast range of geometric diversity in the structure to be moved, it is based on the underlying assumption of the rigidity of the structure. Consequently, such approach can work very well while moving rigid structures under the influence of gravity, they are ineffective while operating in space. Furthermore, robust controls and robust adaptive control strategies for cooperative control of two robot manipulator mounted on non-holonomic mobile robots has been investigated in \cite{LI2008239}. The formulation presented in this article can handle parametric uncertainties, enabling robust control of mobile manipulators in cooperation carrying a common rigid object. However, the authors assumes that an explicit force sensors are mounted on end-effectors of each manipulators. Williams and Khatib \cite{292110} formulated a pioneering concept of `virtual-linkage', a model to characterize internal forces and moments during multi-grasp manipulation. This approach can be leveraged to manipulate objects using multiple mobile manipulators while ensuring accurate control of internal forces on the rigid body of interest. 

Further extension of this works includes Khatib \cite{KHATIB1999175}, where the computation of internal forces are leveraged to devise reconfiguration strategies for the rigid object to be relocated. This article formulates a generalized reconfiguration strategy which enables reconfiguration of the rigid object in constrained 3D environment using multiple mobile manipulators. More recent work such as\cite{7139967} contains robust collaborative manipulation strategies for deformable objects like a bed sheet. However, the proposed study does not require implementing such strategies as the object to be manipulated is rigid. In summary, the existing methods to estimate inertial properties and internal forces on the rigid body, being moved via collaborative manipulation of multiple mobile robots, require an accurate estimate of the forces being applied to the attachment points and resulting motion of the combined system.

TPODS modules consists of four thrusters, which can release a precise stream of compressed air based on the commanded input signals\cite{TPODS_system}. The arrangement of the thrusters enables three Degree of Freedom (DOF) holonomic motion on a planner surface. The TPODS module is equipped with spherical ball transfers, which allows for a near friction-less motion of the module and offsets any gravitational force on the module as the motion of the module is constrained in the plane perpendicular to the gravitational field. For OSAM activities and gaining custody of non-cooperative Resident Space Objects(RSO), TPODS modules have to apply a significant amount of reaction forces on the target body. If not all, majority of the hardware designed to be operated in space goes through an extensive iterative design cycles to achieve extremal strength to weight ratio. This enables lower launch costs of such hardware to space and also keeps the energy required to control attitude of space objects within manageable margins. However, a major consequence of this renders structural elements which are fragile and in need of a special attention and care while being manipulated in space by external servicing modules. Application of external forces on such fragile structure without consideration of their inertial properties can result into catastrophic consequences. 

\section{Objective}
This study aims to devise, implement and verify estimation algorithms based on Kalman filters and nonlinear least squares filters to estimate the inertial properties of TPODS module. The pose of TPODS modules is measured using an accurate high-speed motion capture system. These measurements are fed to the estimation algorithm and the algorithm should estimate the mass and moments of inertia of the body. Finally, the estimated inertial properties are compared to the physical quantities measured to further tune the estimation algorithm to match estimated inertial parameters with their respective ground truth values. 

\begin{table}[b!]
\centering
 \caption{Physical Properties and States, Input and Output of the TPODS plant.}
\parbox{0.3\textwidth}{
\begin{tabular}{|l|l|} 
 \hline
 Property & Value \\ 
 \hline
 Mass & 2.268 kg  \\ 
 Length L & 10 cm \\
 Moment Arm d & 5 cm \\
 Moment of Inertia & \multirow{2}{*}{$\frac{mL^2}{6}$} \\
 along z axis $I_{zz}$ &  \\
 \hline
 \end{tabular}
 \label{table:1}
 }
\hfill  
\parbox{0.65\textwidth}{
\begin{tabular}{|l|l|l|} 
 \hline
 States & Input & Output\\
 \hline
  X position \textbf{x} & Thruster $\bm{T_1}$ & X position \textbf{x}\\ 
  Y position \textbf{y} & Thruster $\bm{T_2}$ & Y position \textbf{y}\\
  Orientation angle $\bm{\Psi}$ & Thruster $\bm{T_3}$ & Orientation angle $\bm{\Psi}$\\
  X speed $\bm{u}$ & Thruster $\bm{T_4}$ & \\
  Y velocity $\bm{v}$ & &\\
  Angular rate $\bm{r}$ & & \\
 \hline
 \end{tabular}
\label{table:2}}
\end{table}

For the analysis of the motion of the module under the influence of various thrust inputs, a Cartesian coordinate system centered at the geometric center of the module and axes parallel to the subsequent side has been considered\cite{TPODS_system}. The X axis is towards the page's right, and the Y axis is towards the top. With these reference axes and the physical properties of the satellite module, as mentioned in Table \ref{table:1}, the motion governing equations can be analyzed considering the state, input, and output variables presented in Table \ref{table:2}. Note that the true values of the mass and moment of inertia of the TPODS module is provided in the table to generate synthetic measurements and subsequent theoretical analysis. In the later section, these quantities will be treated as augmented states and will be estimated with the help of position measurements. 

Considering the planar motion of the module and the arrangement of the thrusters, 3-DOF equations of motion in the body attached reference coordinate system are\cite{nelson1998flight} 
\begin{align} 
\dot{x} &= u\label{eq:EOM1}\\
\dot{y} &= v\\
\dot{\Psi} &= r\\
\dot{u} &= \frac{\Sigma{F_x}}{m}+rv = \frac{\left(T_{2}+T_{3}-T_{1}-T_{4}\right)}{m\sqrt{2}} + rv\\
\dot{v} &= \frac{\Sigma{F_y}}{m}-ru = \frac{\left(T_{1}+T_{2}-T_{3}-T_{4}\right)}{m\sqrt{2}} - ru\\
\dot{r} &= \frac{\Sigma{M_y}}{I_{zz}} = \frac{d\left(T_{1}+T_{3}-T_{2}-T_{4}\right)}{I_{zz}}\label{eq:EOM2}
\end{align}

\section{Estimation Process}
\subsection{Characterization of Thrusters}\label{thruster_char}
\begin{figure}[b!]
    \centering
    \includegraphics[width=0.8\textwidth]{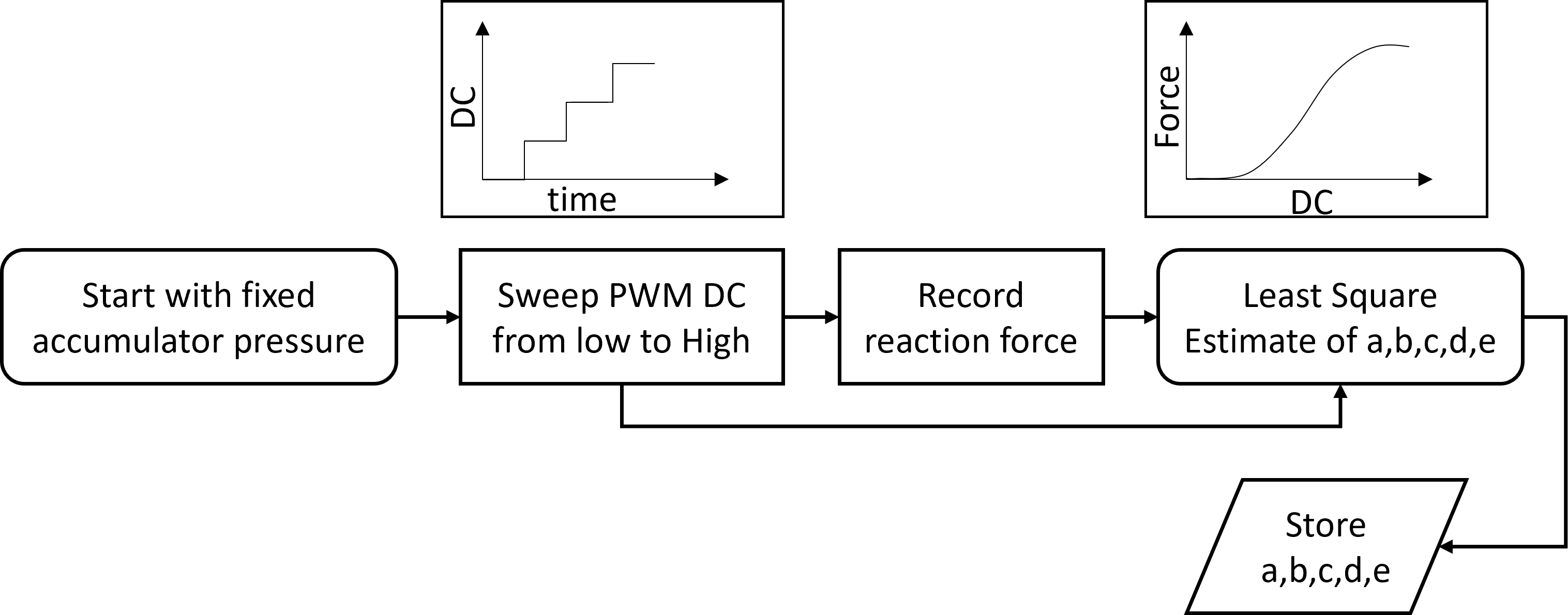}
    \caption{\label{fig:act_estimation}Parameter identification for thrust curve}
\end{figure}

The process of estimation of inertial properties contains primarily two sub-tasks, characterization of thrusters and estimation algorithm. As presented in Figure \ref{fig:act_estimation}, system identification process has been followed to compute the least square estimate of parameters e.g. $a_4, a_3, a_2, a_1$ \& $a_0$ \cite{TPODS_system}. Here, it is assumed that the thrust $T$ from a single nozzle is related to the duty cycle of the PWM control signal, $D$, as per the following equation.
\begin{equation}
    T = a_4D^4+a_3D^3+a_2D^2+a_1D+a_0
\end{equation}

It is important to note that the thrust from nozzles is sensitive to the flow of air through nozzles\cite{osti_110146}. The accumulator pressure and the number of nozzles operating simultaneously dramatically affect the airflow through nozzles. Hence such an identification experiment should be carried out at a fixed accumulator pressure and repeated for various nozzles firing simultaneously. 
\begin{figure}[t!]
\begin{subfigure}[b]{0.49\textwidth}
    \includegraphics[width=\textwidth]{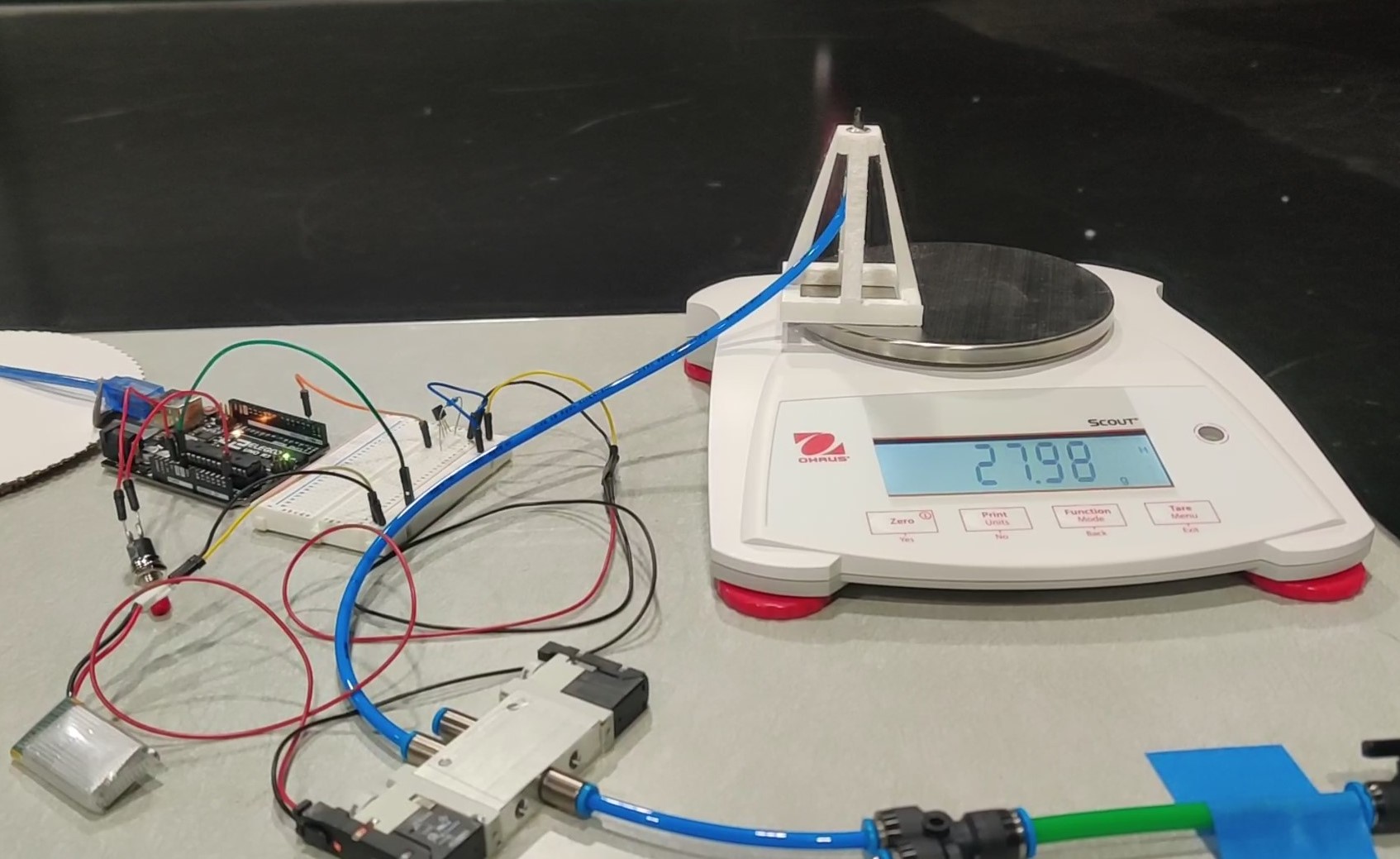}
    \caption{\label{fig:thrust_test}Experimental Setup to measure thrust from nozzle}
\end{subfigure}
\begin{subfigure}[b]{0.49\textwidth}
    \centering
    \includegraphics[width=\textwidth]{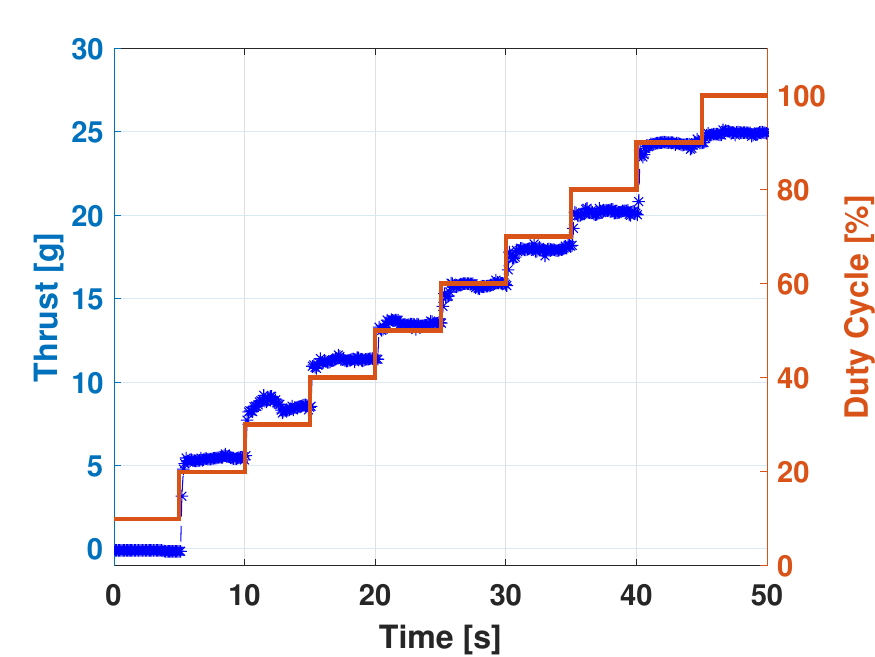}
    \caption{\label{fig:DC_1N}Variation of Duty Cycle and Thrust with time}
\end{subfigure}
\caption{}
\end{figure}

Figure \ref{fig:thrust_test} depicts the experimental setup leveraged to log the variation of thrust from a nozzle in response to the variation in a duty cycle of the solenoid valve. OHAUS Scout electronic weighing scale, along with a USB interface kit are used to measure the thrust with high accuracy and electronically store the thrust data for further analysis. As shown in Figure \ref{fig:DC_1N}, the duty cycle is varied in the step of 10\%, every five seconds. The corresponding variation in the thrust from a single nozzle operating at 60 PSI accumulator pressure is presented in Figure \ref{fig:DC_1N}. The valve is not operational in the duty cycle range below 10\% and above 90\% as the high / low pulse time duration is insufficient to move the internal mechanism \cite{TPODS_system}. Hence for the estimation exercise, only the linear range of the duty cycle (10\%-90\%) is considered.
\begin{align} 
\label{eq:LS1}
h_i &= a_nD_i^n+a_{n-1}D_i^{n-1}+ \dots +a_1D_i+a_0 \\
H &= \begin{bmatrix}
1 & D_1 & D_1^2 & \ldots & D_1^n\\
1 & D_2 & D_2^2 & \ldots & D_2^n\\
\vdots & \vdots & \vdots & \ddots & \vdots \\
1 & D_q & D_q^2 & \ldots & D_q^n\\
\end{bmatrix} \\
z &= \begin{bmatrix} T_1 & T_2 & \dots & T_q\end{bmatrix}^T \\
\hat{x} &= [H^TH]^{-1}H^Tz \label{eq:LS2}
\end{align}

\begin{figure}[h!]
\begin{subfigure}[b]{0.47\textwidth}
    \centering
    \includegraphics[width=\textwidth]{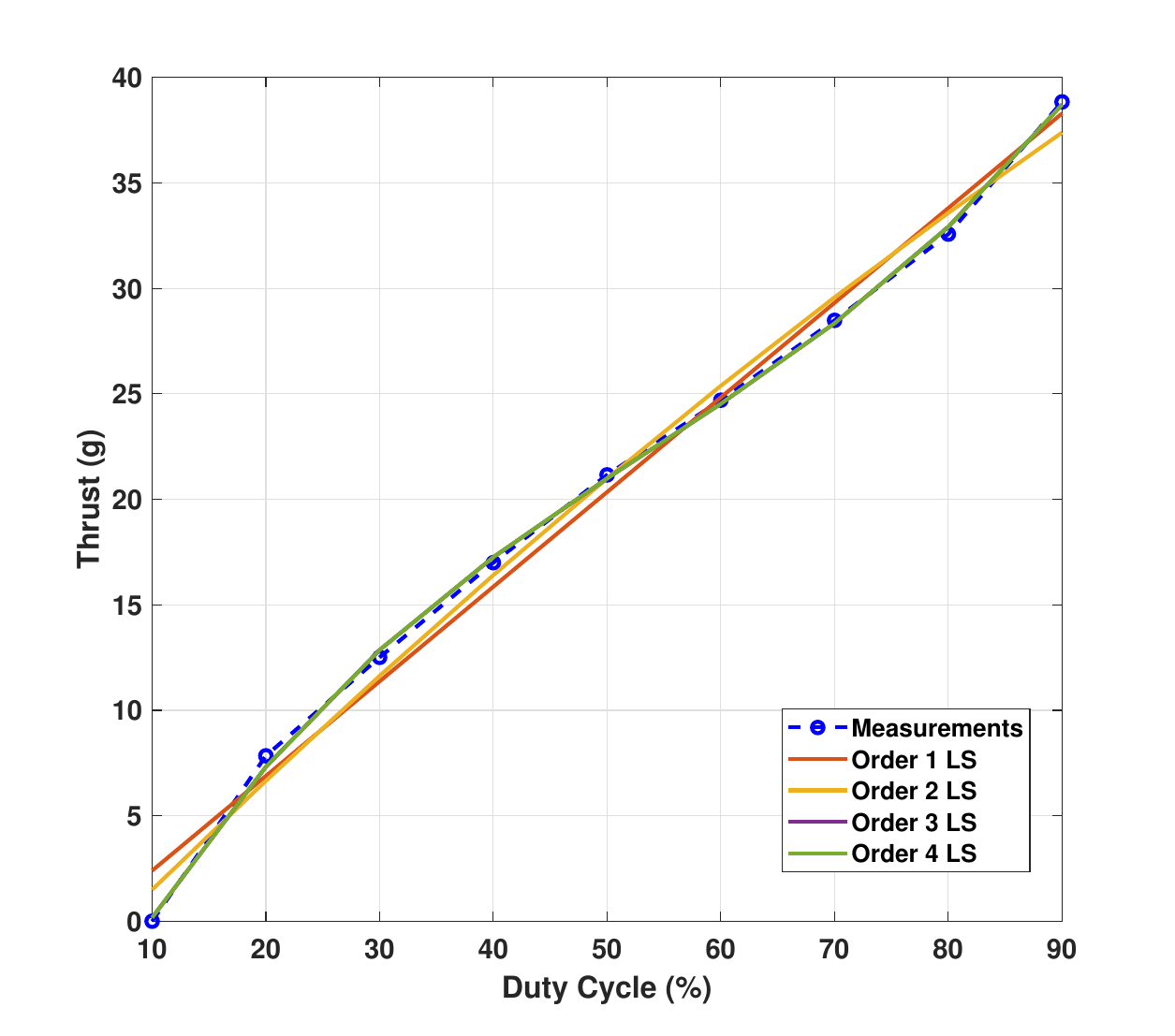}
\end{subfigure}
\begin{subfigure}[b]{0.51\textwidth}
    \centering
    \includegraphics[width=\textwidth]{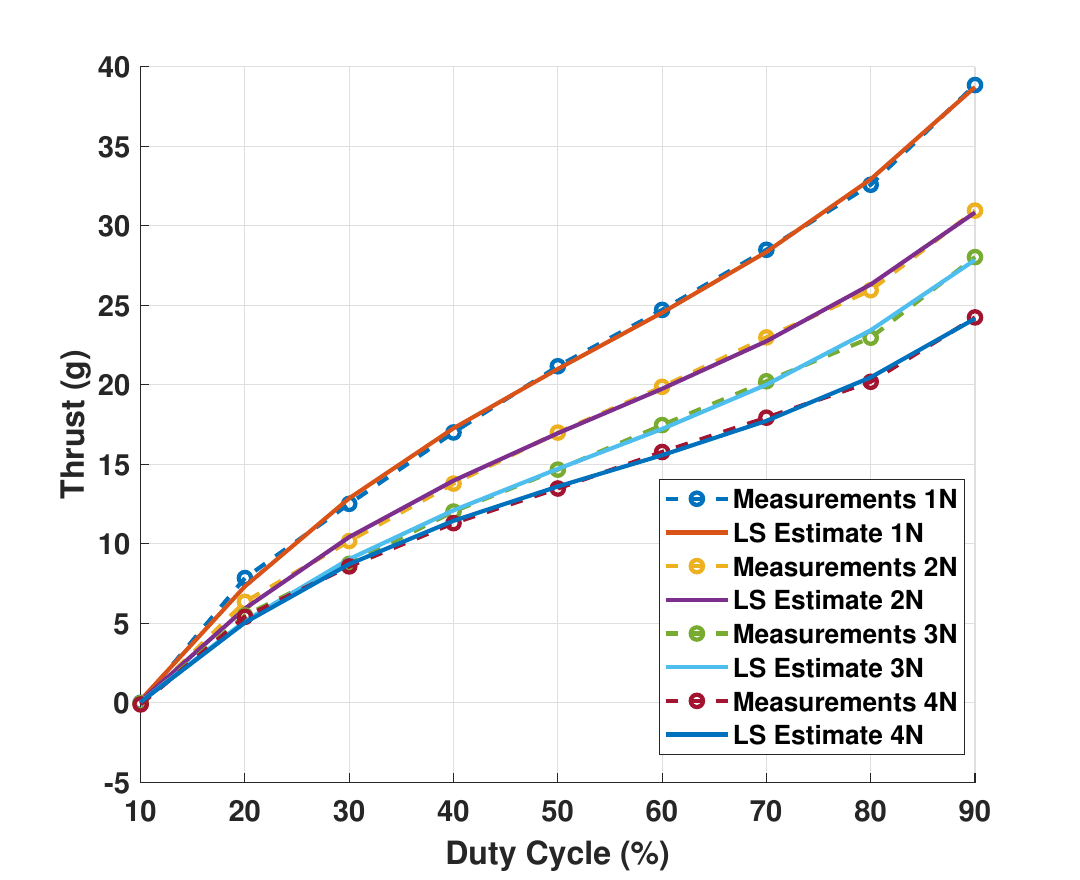}
\end{subfigure}
\caption{\label{fig:LS_final}Comparison of LS estimate of various order and selected third order LS estimate.}
\end{figure}

An average thrust value of the five-second duration is considered to determine the thrust value for each duty cycle variation. With this information, a least square parameter estimation problem was formulated as per Equations \ref{eq:LS1}-\ref{eq:LS2}\cite{crassidis2011optimal}. Here $\hat{x}$ vector consists the least square estimates of parameters, the size of which is determined by the desired order of fit, $n$. With the use of $q$ measurements, we can analyze the effectiveness of the fit for different orders. Such analysis was conducted for the thrust measurements of a single nozzle for orders ranging from one to four. As evident from Figure \ref{fig:LS_final}, the $1^{st}$ and $2^{nd}$ order fits do not accurately describe the actual response. On the contrary, increasing the order beyond three does not significantly improve the fit. Hence, the third-order polynomial fit is selected to estimate the applied thrust based on the given duty cycle. The result of this exercise is presented in Figure \ref{fig:LS_final} for multiple simultaneous nozzle operations and the computed coefficients are given in Table \ref{table:LS_param}.

\begin{table}[h!]
\centering
\caption{Least square parameters for thrust characterization}
\begin{tabular}{|c|c|c|c|c|c|} 
 \hline
 Coefficient & $a_0$ & $a_1$ & $a_2$ & $a_3$ \\ 
 \hline
 One active nozzle & -9.0692 & 1.0439 & -0.0128 & $7.88e^{-5}$ \\ 
 Two active nozzles & -7.4315 & 0.8516 & -0.0104 & $6.35e^{-5}$\\ 
 Three active nozzles & -6.6022 & 0.7555 & -0.0096 & $6.11e^{-5}$ \\ 
 Four active nozzles & -6.6791 & 0.7700 & -0.0104 & $6.34e^{-5}$\\ 
 \hline
 \end{tabular}
 \label{table:LS_param}
\end{table}
 
\subsection{Estimation of inertial properties}
\begin{figure}[h]
    \centering
    \includegraphics[width=\textwidth]{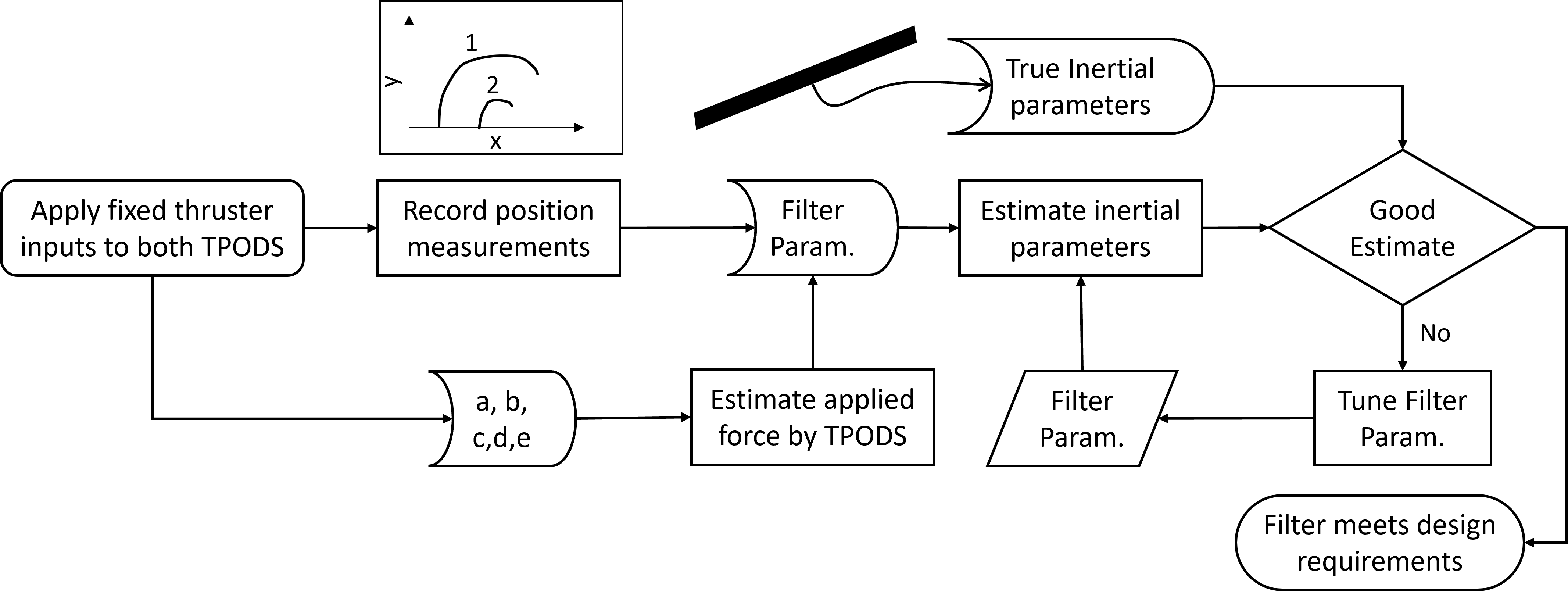}
    \caption{\label{fig:main_estimation}Estimation of inertial parameters using position measurements and applied thrust}
\end{figure}

With experimentally identified actuator model, TPODS can be commanded to move using a predefined thruster firing sequence. This ensures that we have accurate information of applied thrust. The majority of the applied thrust results in the motion of TPODS. However, some portion of the applied thrust also goes towards fighting frictional forces and disturbance due to pneumatic tether. This is an important factor to consider during experimental design and analysis. The frictional losses are relatively constant and can be neglected for the simulation (this assumption is qualifed later based on the experimental data). However, the disturbance due to the pneumatic tether is dependent on the relative position of the TPODS from the center of a support structure. As presented in Figure \ref{fig:main_estimation}, position measurements in response to the commanded input can be recorded. With the prior information about the applied input and parameters that relate commanded PWM signal to thrust applied, the total thrust generated by individual TPODS modules can be computed. Further, the position measurements from the motion capture system combined with applied thrust will be utilized to estimate the inertial properties of the structure. The true value of mass and moments of inertia of the structure provide ground truth data and can be leveraged to assess the performance of the estimation algorithm. Finally, an iterative tuning process needs to be followed to arrive at filter parameters that result in estimation of inertial parameters meeting the design requirements of the estimator. 

\subsection{Estimation using Non-linear Least-squares} \label{sens_analysis}
As summarized in Algorithm \ref{alg:batch_LS}, a non-linear batch least squares algorithm\cite{crassidis2011optimal} is implemented to estimate mass and moments of inertia for a satellite module having a planar motion with 3-DOF. Refining the initial estimated state is repeated untill the correction in the initial state, falls below a provided threshold at the end of the iteration. The system dynamics is augmented by two additional states, representing the parameters to be identified, i.e. mass and moment of inertia\cite{SORENSON198285}. 
\begin{align} 
\label{eq:EOM31}
\dot{m} &= 0 \\
\dot{I_{zz}} &= 0 
\label{eq:EOM32}
\end{align}
The resulting dynamical system of Equations \ref{eq:EOM31}-\ref{eq:EOM32} along with original dynamics are considered for the non-linear least squares state-parameter estimation.
\begin{algorithm}  
\caption{Estimation of inertial properties using batch least squares}\label{alg:batch_LS}
\begin{algorithmic}[1]
\Require a set of data containing applied thrust input $T(t)$ and pose state $X(t)$ with associated measurement noise covariance $P_{vv}$ and threshold $thr$ to exit the iterative estimation process 
\State $ x_{0}^*\gets $ Reference state, $ \delta \Bar{x}_{0}\gets $ A prior estimate, $ \Bar{P}_{xx,0}\gets $ A prior covariance
\While{$max(abs(\delta \hat{x}_0)) > thr$}
    \State Initialize the accumulation loop with $l=1$, $t_{i-1}=t_0$, $x^*(t_{l-1})=x^*_0$, $\lambda = \Bar{P}_{xx,0}^{-1}\delta \Bar{x}_{0}$, $\Lambda = \Bar{P}_{xx,0}^{-1}$ and $\Phi(t_{l-1},t_0) = I$
    \For{$l=2$ to $t_f$}
    \State Parse the $l^{th}$ measurement to compute $t_l$, $z_l$, $P_{vv,l}$
    \State Integrate the reference state and state transition matrix from $t_{l-1}$ to $t_l$ \Return $x_l^*$ and $\Phi(t_l,t_0)$
    \State Accumulate the current observation by updating $\lambda$ and $\Lambda$
    \EndFor
    \State Solve the normal equations $\Lambda \delta \hat{x}_0 = \lambda$ to find the estimate and compute the covariance $P_{xx,0} = \Lambda^{-1}$
    \State Return to Step 3 with $x_0^* = x_0^* + \delta \hat{x}_0$, $\delta \Bar{x}_0 = \delta \Bar{x}_0 - \delta \hat{}{x}_0$
\EndWhile
\end{algorithmic}
\end{algorithm}

\begin{figure}[h!]
\begin{subfigure}[b]{0.52\textwidth}
    \centering
    \includegraphics[width=\textwidth]{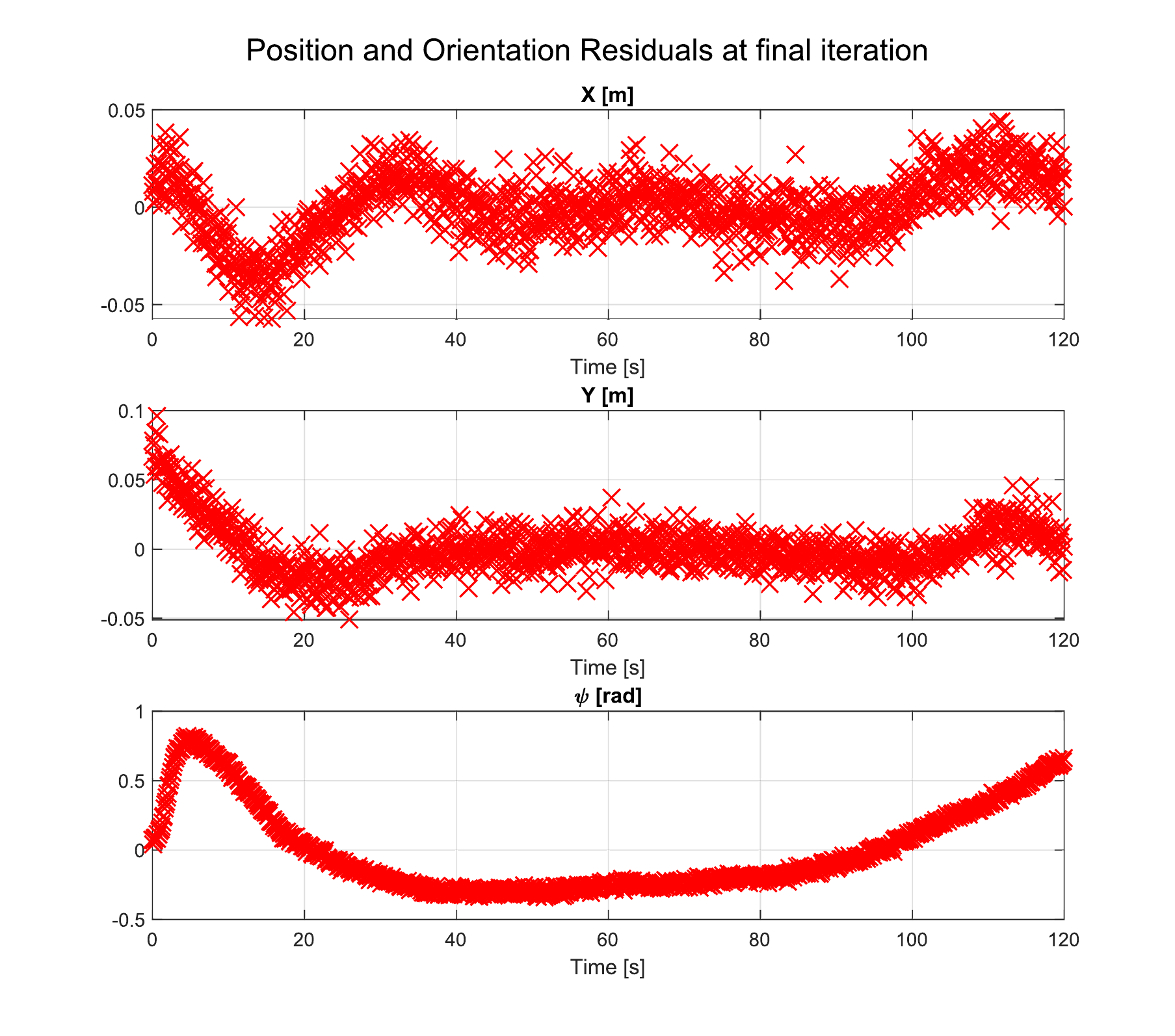}
    \caption{\label{fig:LS_param_1}Least Square residuals for orbit follower}
\end{subfigure}
\begin{subfigure}[b]{0.47\textwidth}
    \centering
    \includegraphics[width=\textwidth]{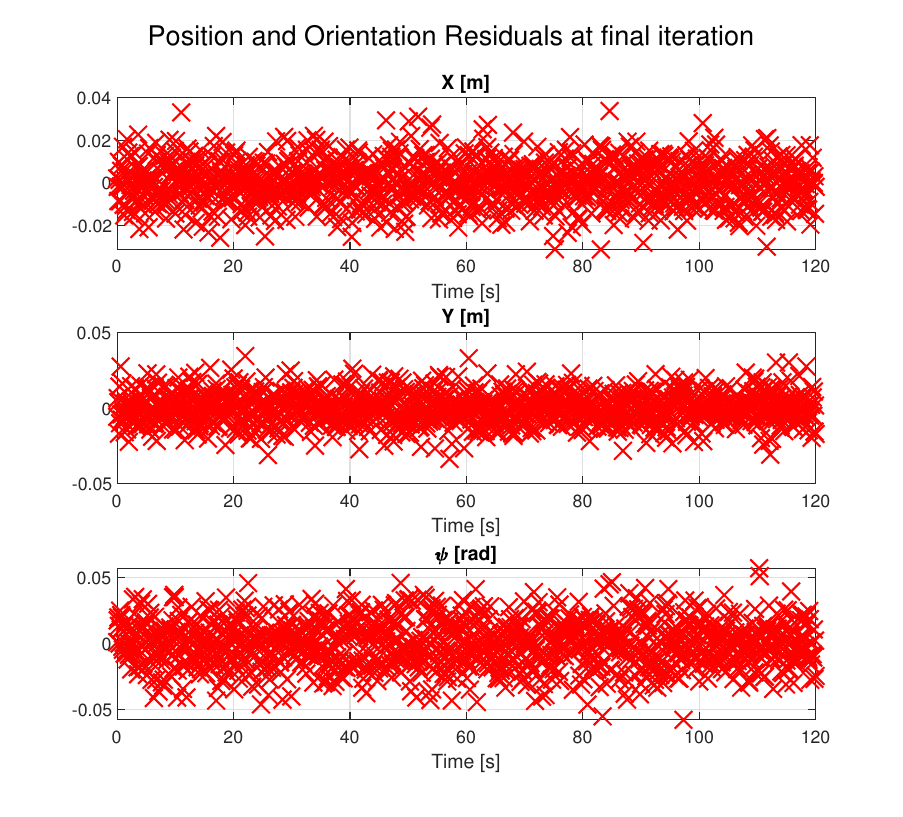}
    \caption{\label{fig:LS_param_2}Least Square residuals for informed input based on the sensitivity analysis}
\end{subfigure}
\end{figure}

For the preliminary analysis, a phase-sifted sine wave of magnitude 0.1m and a period of 60s is given as the desired position setpoints to the closed-loop controller. The resulting motion and control input history is logged. Synthetic pose data is generated by injecting additive Gaussian noise with a standard deviation of 1cm in position and 1 degree in orientation to the true pose. Figure \ref{fig:LS_param_1} presents the residuals at the end of the final iteration when the threshold was set as $1e^{-8}$. It is evident from Figure \ref{fig:LS_param_1} that the residuals are not symmetrically distributed across the zero mean, highlighting the fact that the estimated parameters are not satisfactorily close to their true values\cite{NONLS}. This is further supported by the analysis presented in Table \ref{table:4}, where the estimated parameters for the orbit follower case have large divergence from true values.
\begin{table}[h!]
\centering
\caption{Least square estimated of inertial parameters}
\begin{tabular}{c|c|c|c|c|} 
 \cline{2-5}
  & \multicolumn{2}{|c|}{Orbit follower} & \multicolumn{2}{|c|}{Phase shifted sine}  \\ 
 \cline{2-5}
  & Mass [$kg$] & Moment of Inertia [$kg$ $m^2$]& Mass [$kg$] &  Moment of Inertia [$kg$ $m^2$] \\
 \hline
  \multicolumn{1}{|c|}{Initial} & 1 & 0.005 & 1 & 0.005 \\
  \hline
  \multicolumn{1}{|c|}{True} & 2.268 & 0.00378 & 2.268 & 0.00378\\
  \hline
  \multicolumn{1}{|c|}{Estimated} & 2.317 & 0.0061 & 2.2661 & 0.00378\\
  \hline
  \multicolumn{1}{|c|}{Absolute error} & 0.049 & 0.0023 & 0.0018 & $1.066e^{-6}$\\
  \hline
  \multicolumn{1}{|c|}{Relative error} & 2.16 \% & 61.38 \% & 0.078 \% & 0.0028 \% \\
  \hline
 \end{tabular}
 \label{table:4}
\end{table}

\begin{figure}[h!]
\begin{subfigure}[b]{0.49\textwidth}
    \centering
    \includegraphics[width=\textwidth]{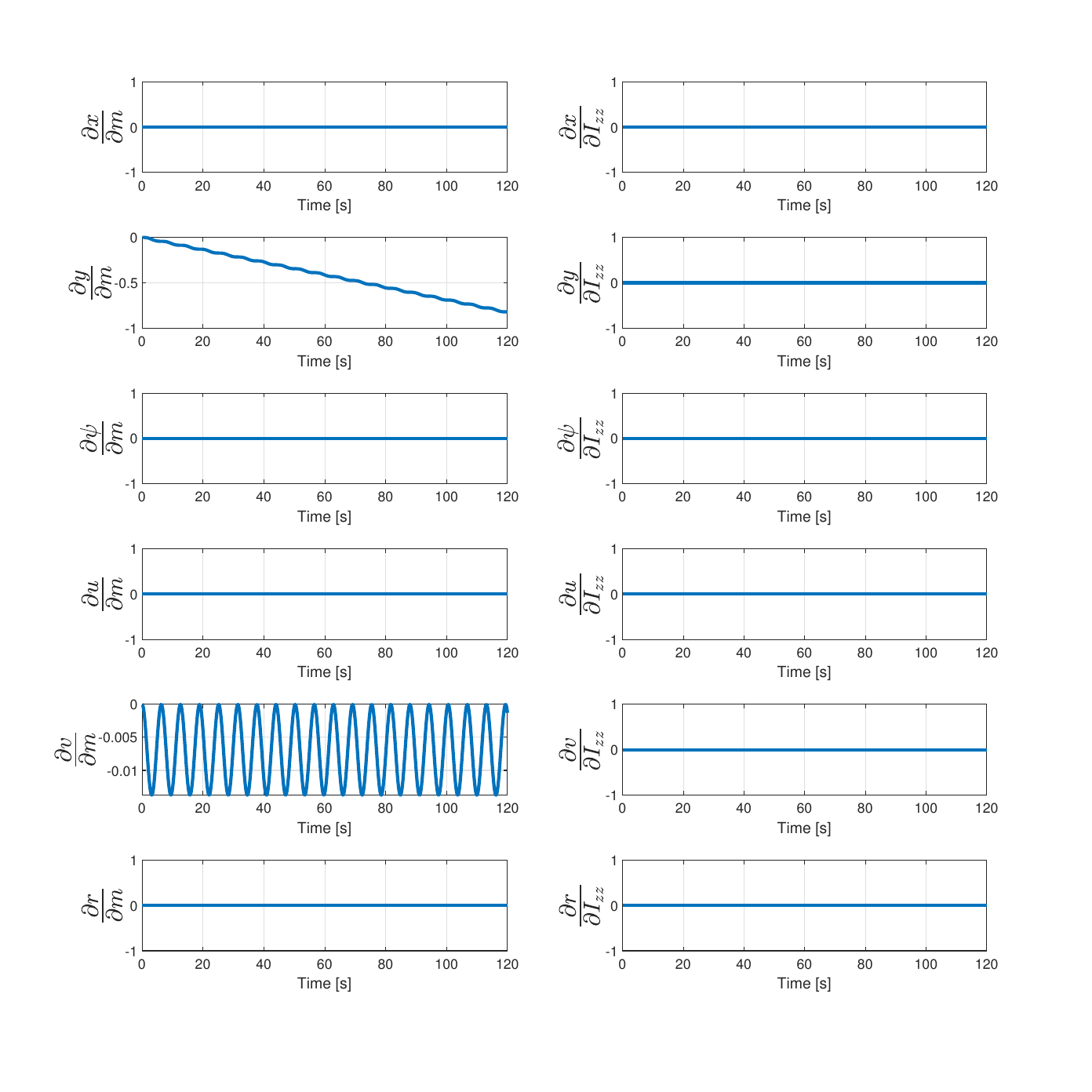}
    \caption{\label{fig:sens_1}}
\end{subfigure}
\begin{subfigure}[b]{0.49\textwidth}
    \centering
    \includegraphics[width=\textwidth]{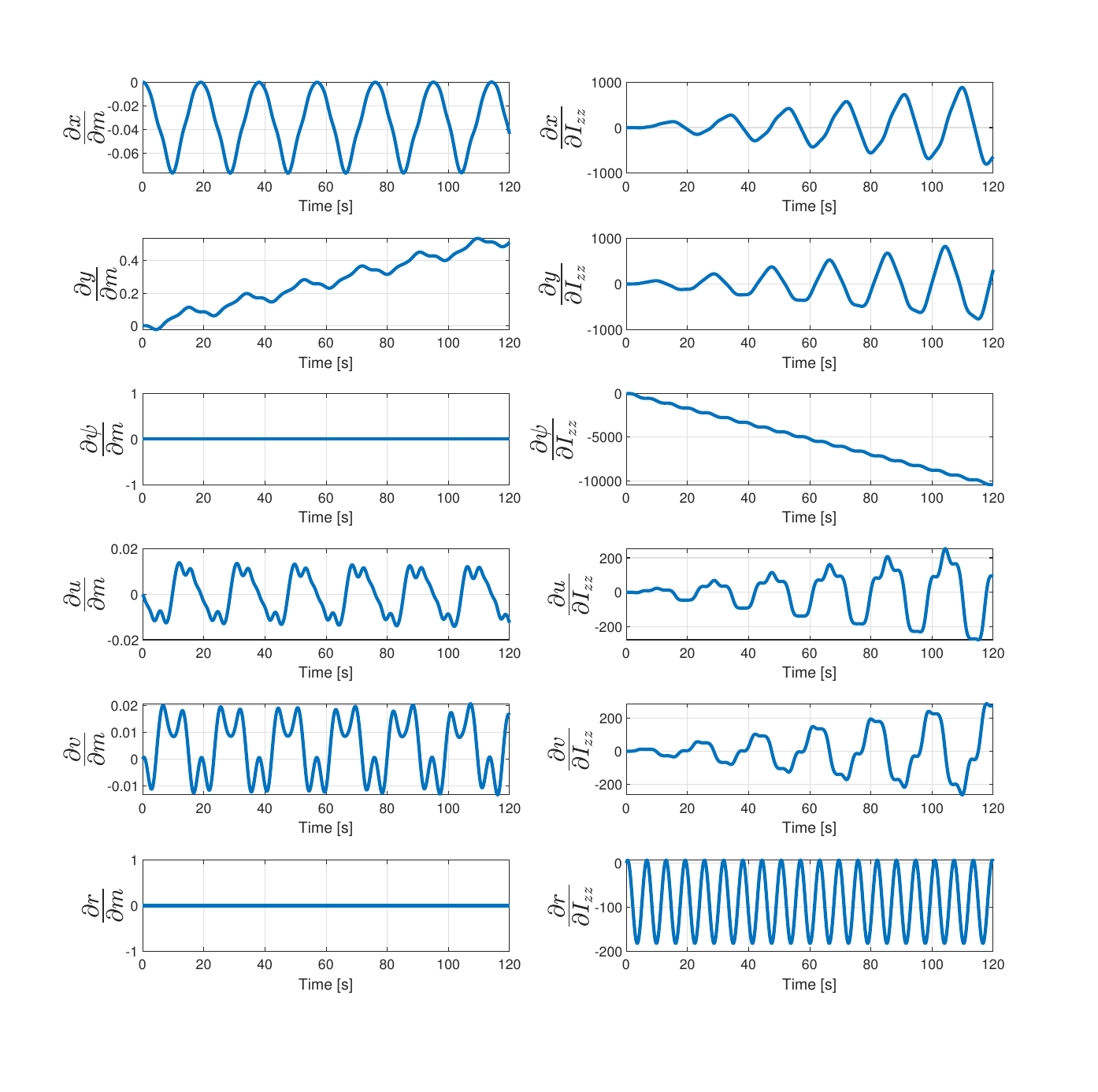}
    \caption{\label{fig:sens_2}}
\end{subfigure}
\caption{\label{fig:sens_1_2}State and Output Sensitivity for simple and informed input}
\end{figure}

One of the probable reason for the divergence of estimated values from their true values is lack of observability of the estimated states in the sate history. The trajectory followed by the module is dependent on the nature of the applied input. Hence, it is vital to identify an efficient and condusice input sequence that results in accurate estimation of inertial parameters. The sensitivity analysis is further employed to rectify the parameter divergence issues and identify appropriate input sequence \cite{doi:10.1021/acs.iecr.0c03793}. Equation \ref{eq:sen1} depicts the dynamical system by combination of inputs $u$, states $x$ and unknown parameters $\theta$. 
\begin{align} \label{eq:sen1}
\dot{x} &= F(x,u,\theta) \\
y &= H(x,\theta) 
\end{align}
Based on the system behavior sensitivity partials ${S}_{x,\theta}$ can be calculated for a given input sequence by numerical integration of $\dot{S}_{x,\theta}$.
\begin{align}
S_{x,\theta} &= \frac{\partial x}{\partial \theta} \\
\dot{S}_{x,\theta} &= \frac{\partial F }{\partial x}S_{x,\theta} + \frac{\partial F }{\partial \theta}
\end{align}
Finally, the time history of ${S}_{x,\theta}$ can be leveraged to calculate the output sensitivity partial ${S}_{y,\theta}$ as per Equations \ref{eq:sen2}. 

\begin{align}
S_{y,\theta} &= \frac{\partial H}{\partial x}S_{x,\theta} + \frac{\partial H}{\partial \theta}
\label{eq:sen2}
\end{align}
These partials indicate a degree of observability for parametric change in the system output. An input signal that results in oscillatory y velocity only is presented in Figure \ref{fig:sens_1}. It can be inferred from the plots that with such an input signal, only mass can be estimated with high accuracy. However, the output is not sufficient to predict the moment of inertia. To find an input sequence that can result in output history, showcasing excellent observability for both parameters, the persistence of excitation property of a signal is leveraged\cite{ioannou2013robust}. 

One such input sequence is phase shifted sine wave of magnitude 0.025, frequency of 1 rad/sec and shifted by $\pi/4$ in phase for each input i.e. first input has zero phase, the second input has phase $\pi/4$ and so on. It is evident from Figure \ref{fig:sens_2} that the phase-shifted sine wave inputs produce an output having a significant sensitivity for each parameter. A similar process of generating synthetic data and batch least square estimates is followed for this input sequence and the results are presented in Figure \ref{fig:LS_param_2} and Table \ref{table:4}. The residuals for this case are evenly spaced around the center line, indicating that they have a zero mean, essentially recovering the Gaussian measurement noise\cite{NONLS}. The numeric analysis presented in Table \ref{table:4} further supports this fact, as the robust parameter convergence is achieved using the input sequence inferred by the stability analysis. 

\subsection{Estimation using Extended Kalman Filter}
\begin{figure}[b!]
\begin{subfigure}[b]{\textwidth}
    \centering
    \includegraphics[width=\textwidth]{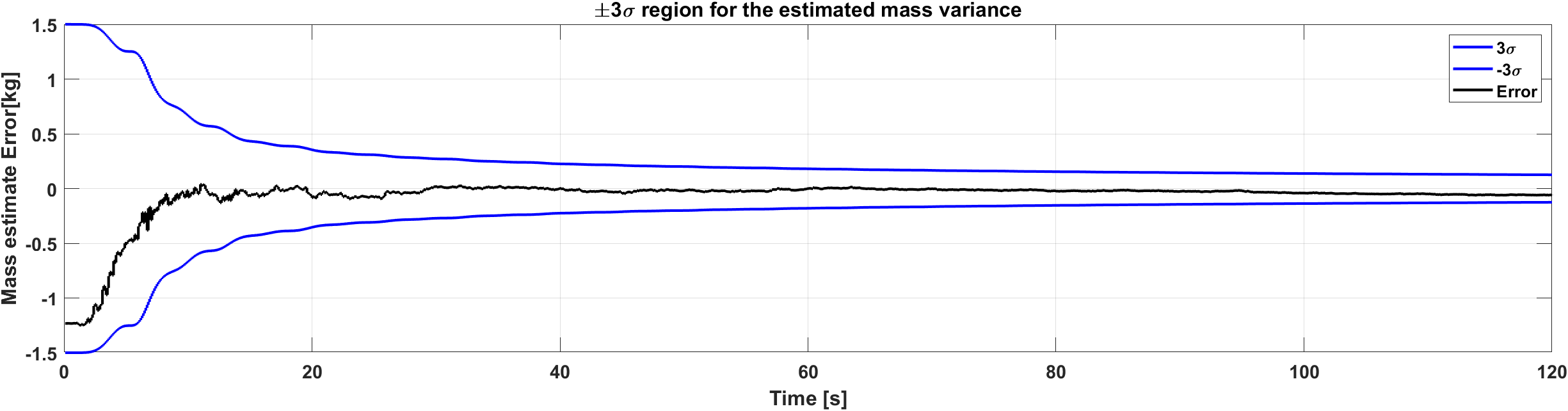}
\end{subfigure}
\begin{subfigure}[b]{\textwidth}
    \centering
    \includegraphics[width=\textwidth]{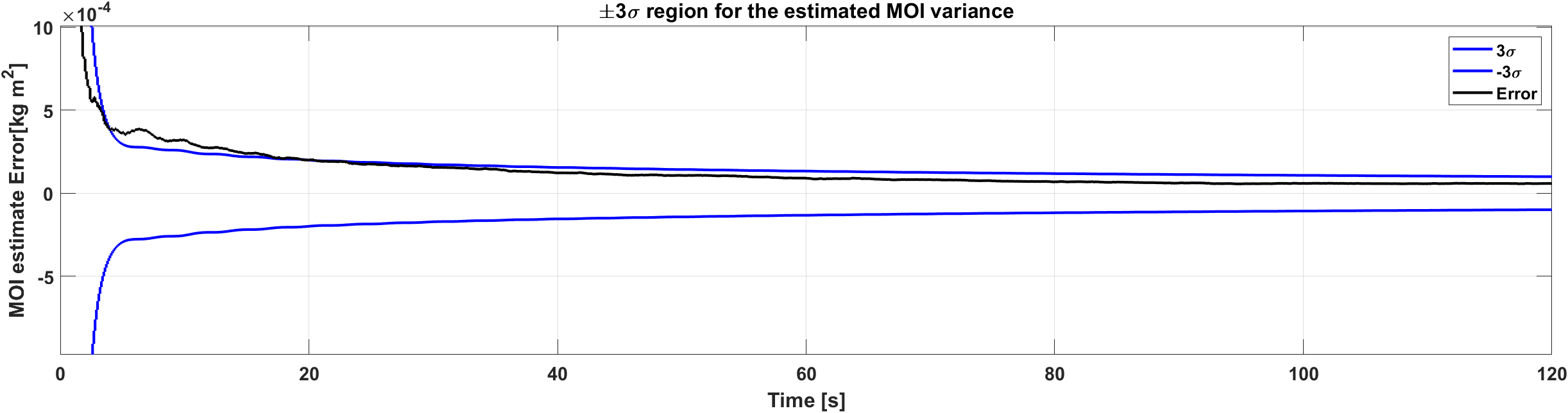}
\end{subfigure}
\caption{\label{fig:EKF}Variance and error history of mass and MOI estimates}
\end{figure}

Based on the dynamical system described by Equations \ref{eq:EOM31}-\ref{eq:EOM32}, an extended Kalman filter(EKF) algorithm is implemented for an online estimation of inertial parameters\cite{crassidis2011optimal}. For the preliminary analysis, it is considered that there is no process noise present. However, upon inspection of the estimation error and standard deviation, it is observed that the estimation error for mass was persistently leaving the $3\sigma$ region. Hence, process noise of $1e^{-4} kg^2/Hz$ is injected in the governing equation of mass i.e. $\dot m = 0$. In addition, the initial conditions are sampled from a Gaussian distribution having mean of [0 m 0 m 0 deg 0 m/s 0 m/s 0 deg/s 1 kg $5^{-3}$ kg $m^2$] and standard deviation of [0.1m 0.1m 5deg 0.1m/s 0.1m/s 2deg/s 1kg $2^{-3}kg$ $m^2$] respectively. The measurement variance of [0.01$^2m^2$ 0.01$^2m^2$ $1deg^2$] is considered in accordance with the accuracy of the motion capture system. The filter performance is evaluated using the same synthetic data generated in the previous section and the results are presented in Figure \ref{fig:EKF}. It can be observed that the filter performance is satisfactory and overall, it performs well in estimating mass and moments of inertia. The variances converge quickly to a lower value and keep getting lower as time progresses. The Root Mean Square Error is 0.2267 kg for mass estimation and $2.85\times10^{-4}kgm^2$ for the moment of inertia estimation. However, it is also important to acknowledge the fact that the estimation error for moment of inertia leaves the $3\sigma$ region for a considerable amount of time. This can be rectified by injecting more process noise, resulting in a lesser accurate estimate. 

\begin{figure}[t!]
\begin{subfigure}[b]{0.49\textwidth}
    \centering
    \includegraphics[width=\textwidth]{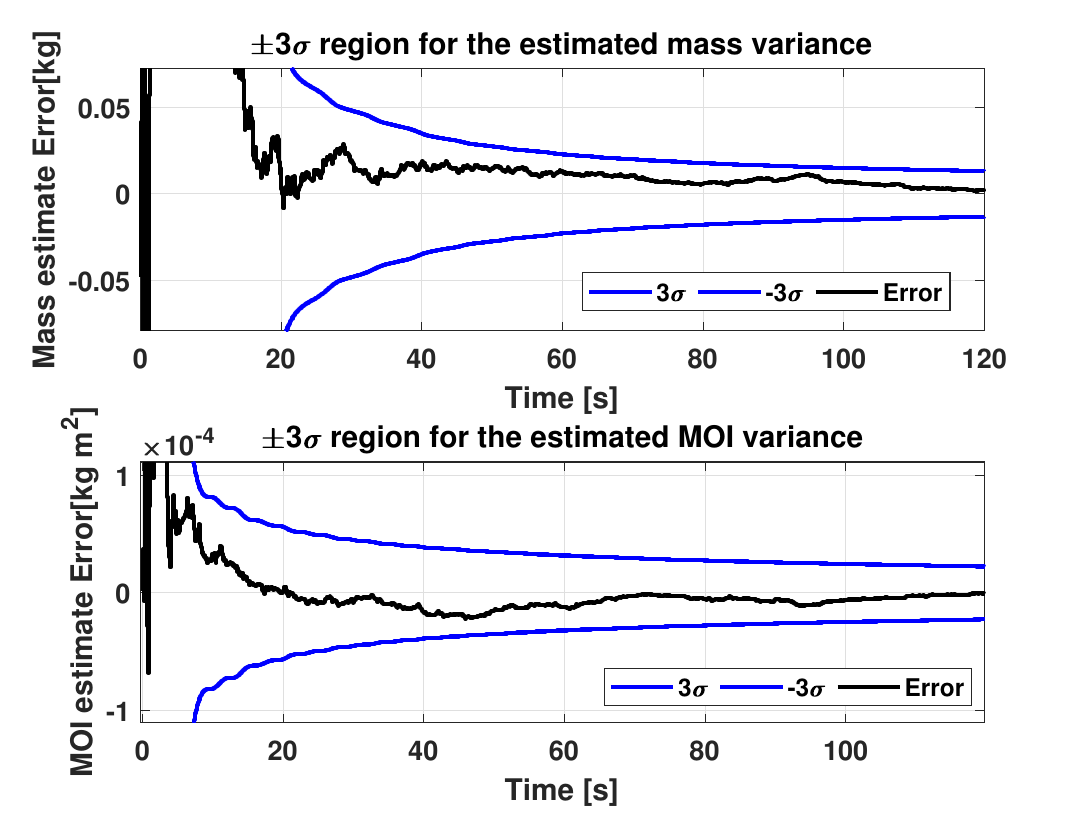}
\end{subfigure}
\begin{subfigure}[b]{0.49\textwidth}
    \centering
    \includegraphics[width=\textwidth]{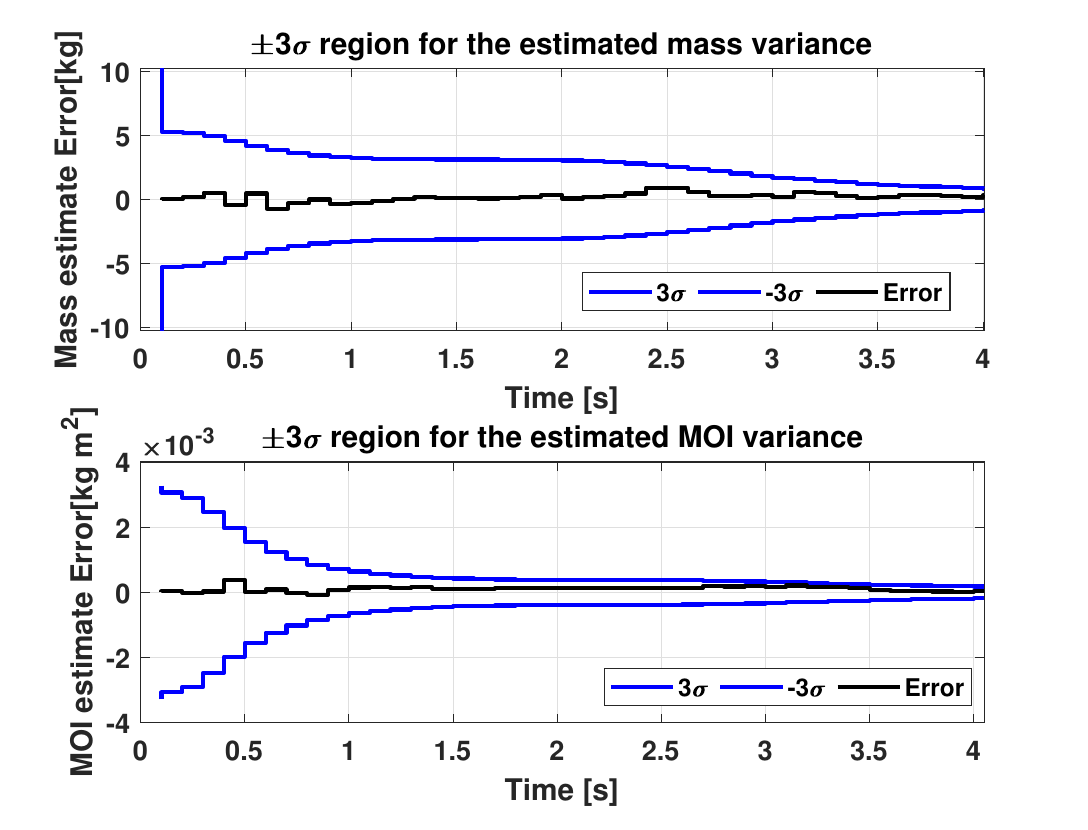}
\end{subfigure}
\caption{\label{fig:EKF_LS}Variance and error history of mass and MOI estimates}
\end{figure}
One of the probable reasons for departure from the $3\sigma$ region and the bias in estimated parameters is an absence of cross-covariance in the initial covariance guess. The fact that no unique set of parameters satisfies the dynamics equations resulting in the provided measurements also poses additional challenges pertaining to the convergence of estimated parameters to their true values. The batch least squares estimate algorithm is utilized to compute a good initial covariance matrix to rectify this issue. From the total available data of 120 seconds, the first 20 seconds of data is processed using the batch least squares process discussed in the previous section and the resulting estimated initial state and covariance estimates are used for the initial guess for the EKF. The initial mean resulting from such an approach is very close to the respective true values. However, the initial covariance has significant diagonal elements, depicting low confidence in the estimated mean values. The EKF further processes the whole data of 120 seconds and produces estimated mean within 0.1\% of the true mass and 0.01\% of the true moment of inertia values. The resulting covariance also has standard deviations of within 0.2 \% of the true mass and 0.2 \% of the true moment of inertia values. The resulting estimation errors, along with $3\sigma$ regions for mass and moments of inertia, are presented in Figure \ref{fig:EKF_LS}, which testifies the effectiveness of the combined LS and EKF approach for estimating inertial properties. The Root Mean Square Error is 0.11 kg for mass estimation and $1.74\times10^{-5}kgm^2$ for the moment of inertia estimation, which is significantly better than the EKF alone.

\section{Robustness Analysis}
\begin{figure}[t!]
\centering
\begin{subfigure}[b]{0.49\textwidth}
    \includegraphics[width=\textwidth]{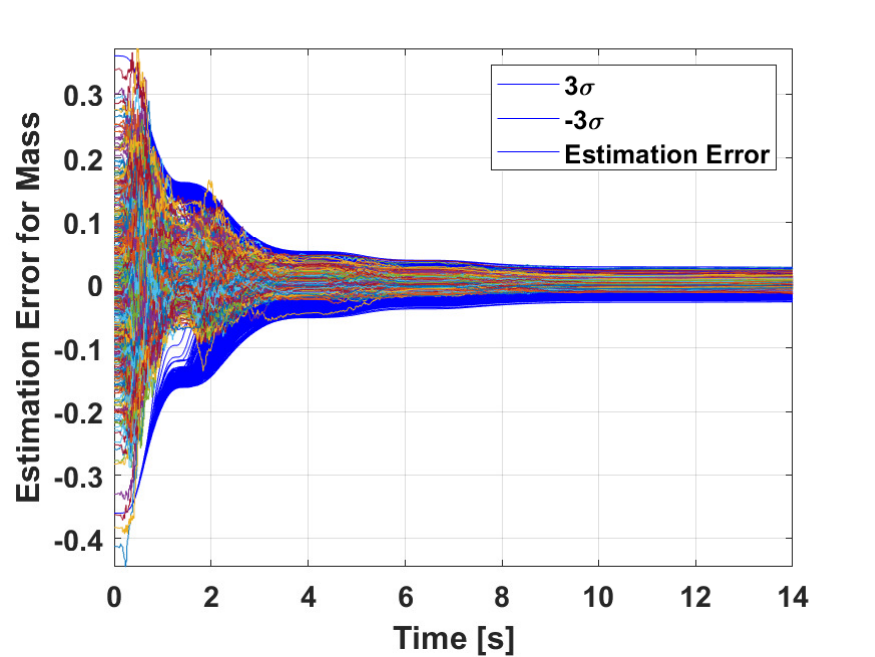}
    \caption{Variance and error history of mass}
\end{subfigure}
\hfill
\begin{subfigure}[b]{0.49\textwidth}
    \centering
    \includegraphics[width=\textwidth]{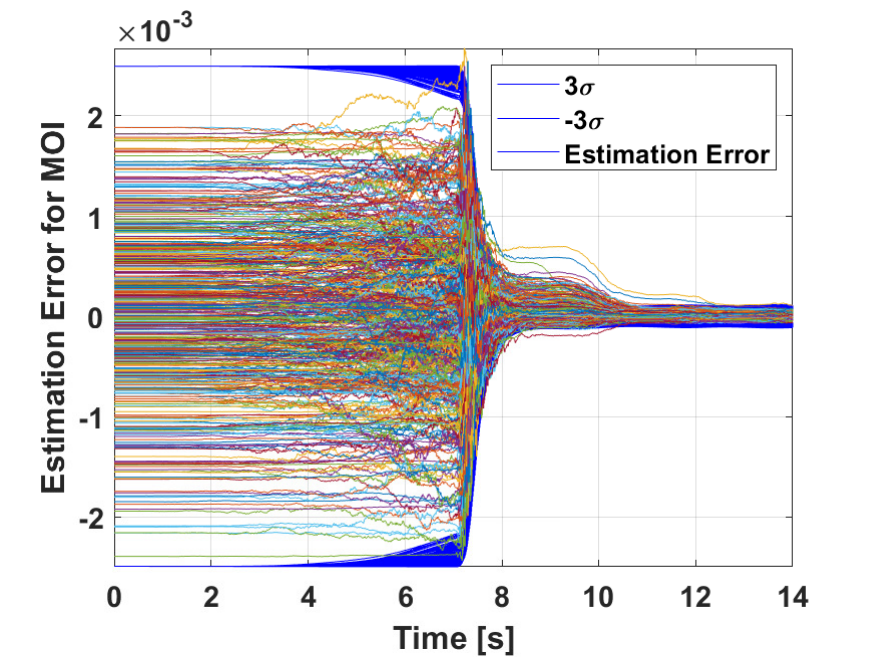}
    \caption{Variance and error history of MOI}
\end{subfigure}
\caption{\label{fig:MC}Monte Carlo analysis}
\end{figure}

The filter designed in the earlier section is further analysed for the robustness using Monte Carlo analysis of 500 distinct simulation scenarios. For the analysis, the inertial parameters are varied till $\pm50\%$ of their true values. Filter is initialized at their true values for all of the simulation scenarios and the true trajectory is computed with adequate process noise. Further, the synthetic measurements are recorded according to the measurement model described in the earlier section. In each case, the estimation algorithm utilized the available measurements and the dynamic model to infer an educated guess of the inertial parameters of the TPODS module. As show in Figure \ref{fig:MC}, the estimation error for mass and MOI both remains withing the $\pm3\sigma$ regions. The standard deviation of estimated quantities also reduces gradually as more reliable information is available. This exhibits the effectiveness and robustness of the estimation algorithm.  

\begin{figure}[b!]
\centering
\begin{subfigure}[b]{0.49\textwidth}
    \includegraphics[width=\textwidth]{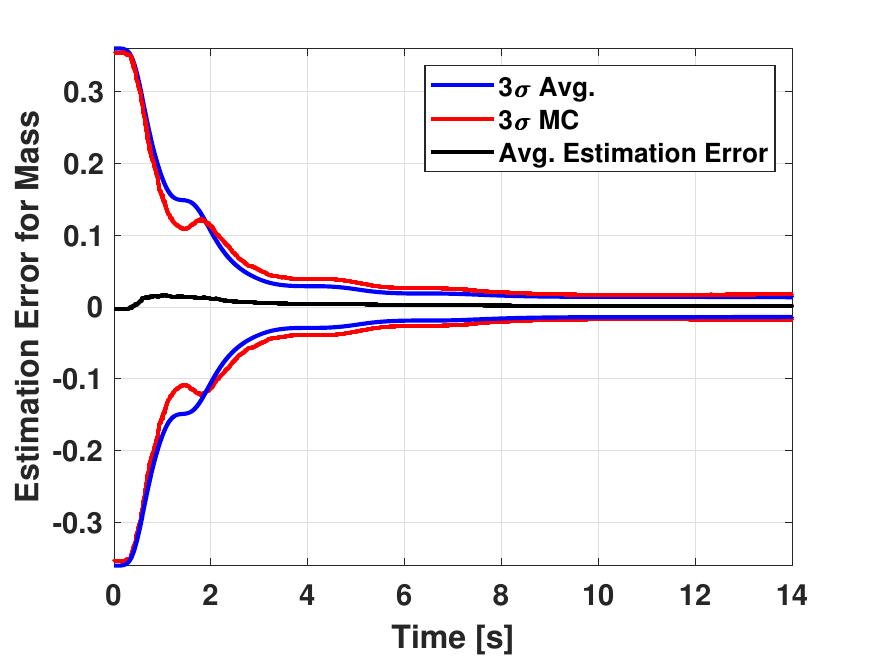}
    \caption{Average variance and error history of mass}
\end{subfigure}
\hfill
\begin{subfigure}[b]{0.49\textwidth}
    \centering
    \includegraphics[width=\textwidth]{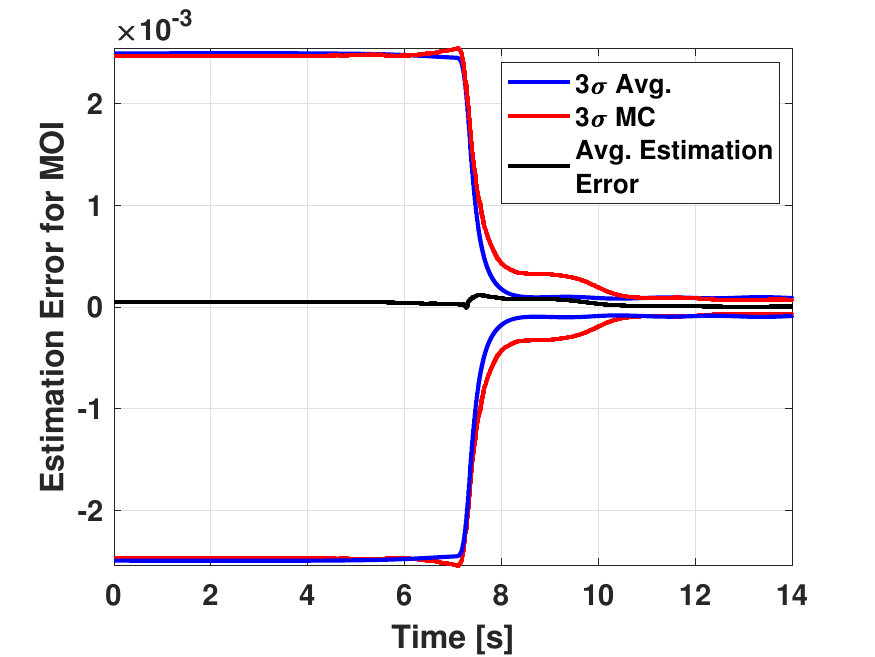}
    \caption{Average variance and error history of MOI}
\end{subfigure}
\caption{\label{fig:MC_overall}Average quantities from Monte Carlo analysis}
\end{figure}

A further analysis of average standard deviation, average estimation error and spread of errors are a particular time instance for all simulation scenarios reveals some of the fundamental properties of the filter. As shown in Figure \ref{fig:MC_overall}, the average estimation error for mass and MOI both remains close to zero, revealing that the estimation algorithm is unbiased. In addition to that observing the average covariance and spread of errors (also referred to as sample covariance), we can conclude that the estimation algorithm is statistically consistent. The partial inconsistency between sample covariance and average covariance for MOI can be potentially due to a sudden availability of accurate measurements with high observability for MOI. Around seven seconds, the module is applied with an excitation that results in pure rotational motion. However, the time duration beofre this is dominated by input sequence that only results into translation motion. Hence, this sudden drop in covariance for MOI is potentially causing the departure of error from $\pm3\sigma$ region. This behavior can be corrected using the well known underweighting modification to the update step\cite{zanetti2010underweighting}. 

\section{Experimental Analysis}
The estimation algorithm developed in the earlier section is further validated using an experiment designed to log the pose history of the TPODS module under the influence of a predefined input sequence. The input sequence is selected in accordance with the observability analysis conducted in the earlier section. The pose measurements include planar position coordinates and a rotation with respect to the axis perpendicular to the plane of motion. The following subsections provide comprehensive details about a design revision of TPODS module, measurement of true inertial parameters, experimental setup and subsequent analysis. 

\subsection{Revised TPODS Module}
\begin{figure}[b!]
\centering
\begin{subfigure}[b]{0.49\textwidth}
    \includegraphics[width=\textwidth]{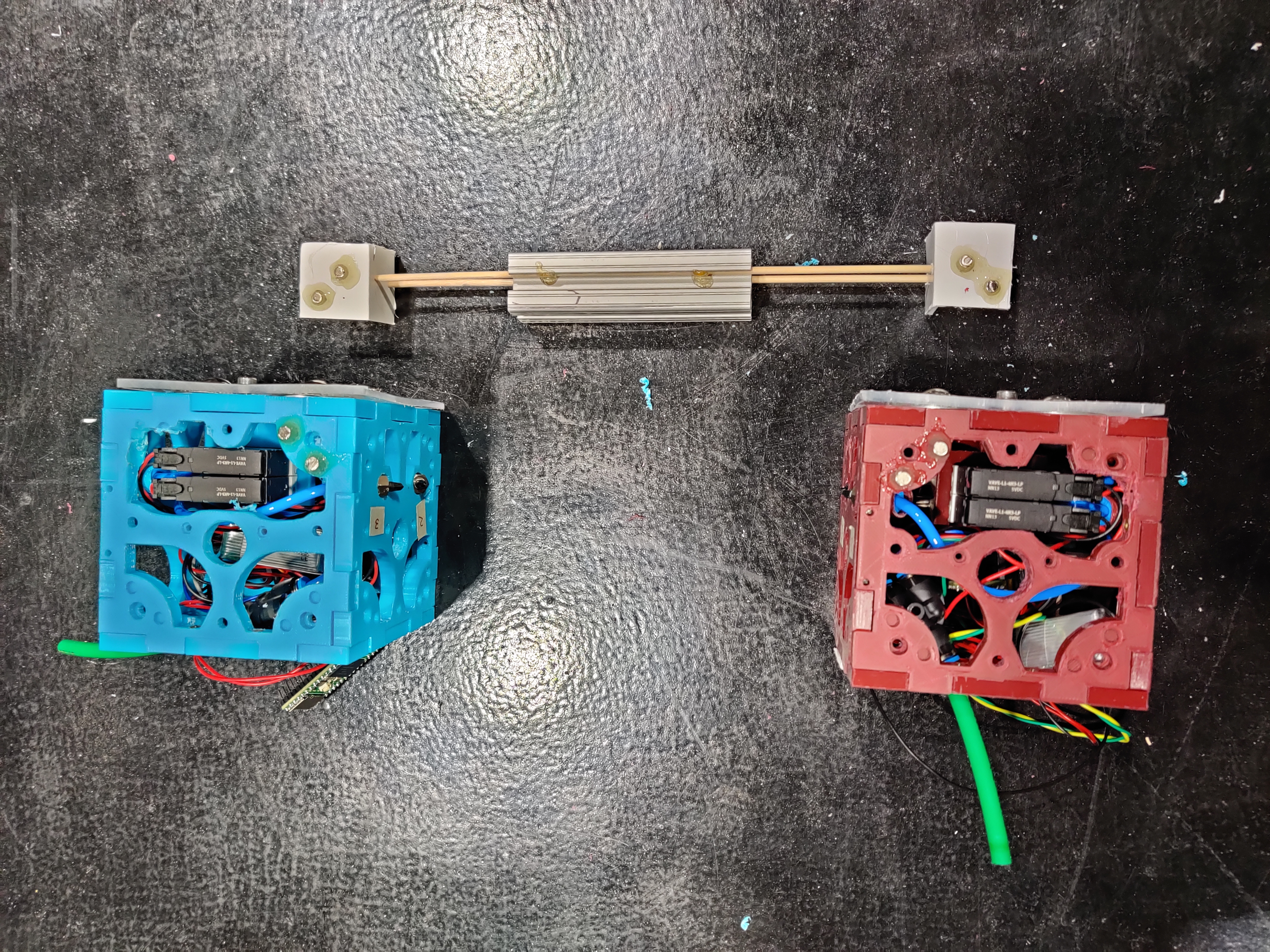}
    \caption{\label{fig:TPODS_MKIII}Revised TPODS modules with external body to be moved}
\end{subfigure}
\hfill
\begin{subfigure}[b]{0.49\textwidth}
    \centering
    \includegraphics[width=0.8\textwidth]{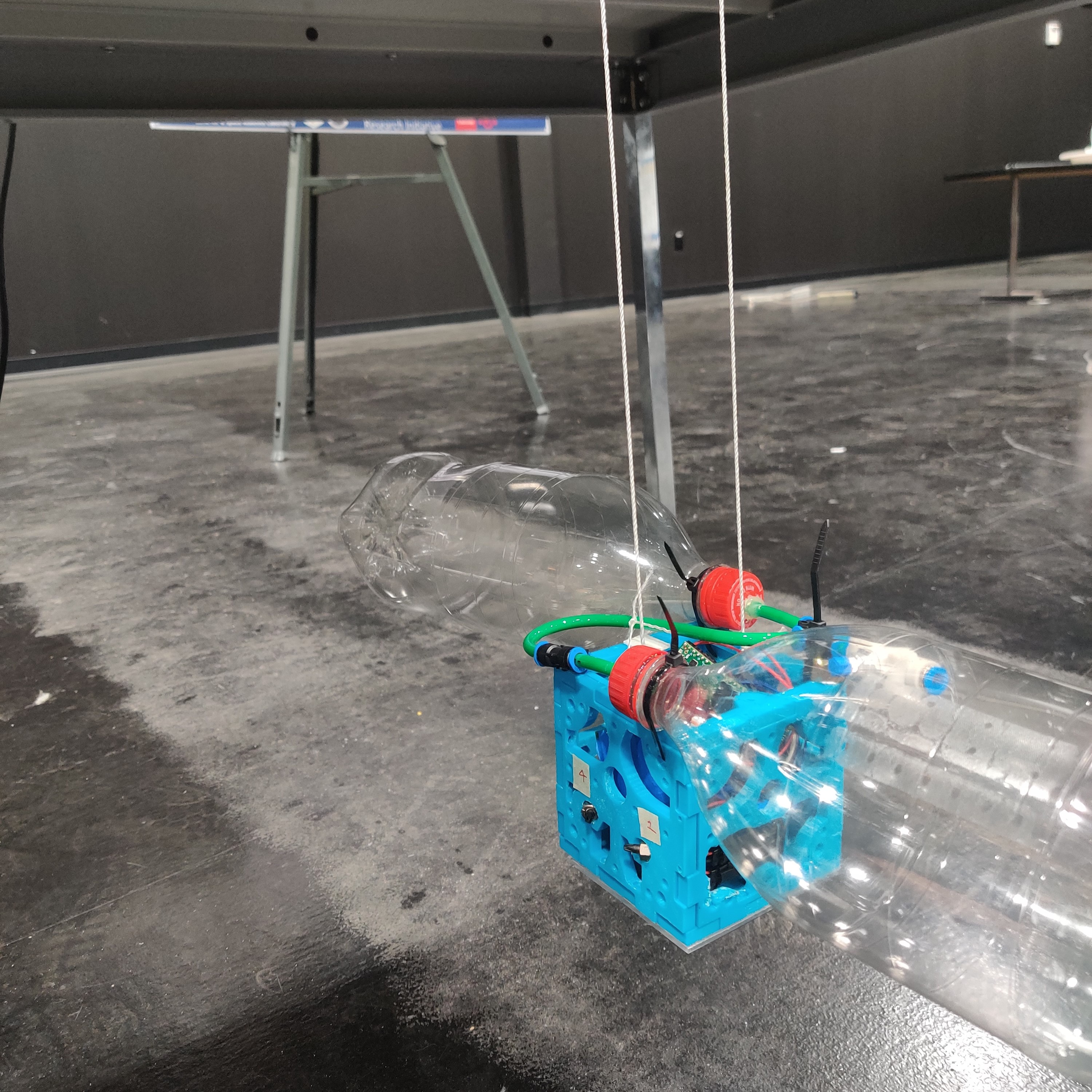}
    \caption{\label{fig:TPODS_MOI}Measurement of true moment of inertia}
\end{subfigure}
\caption{}
\end{figure}

A few significant design improvements have been carried out on the TPODS module to enable the intended end goal of RPOD, satellite servicing and detumbling of RSOs. Compared to the MK-II module depicted in \cite{TPODS_system}, the MK-III modules presented in Figure \ref{fig:TPODS_MKIII} boast a compact design. One of the first notable changes is the incorporation of solenoid control valves inside the volume of the satellite module instead of ad-hoc mounting on the outer walls. In addition to the apparent improvement in appearance, this also helps improve the satellite module's stability. Control valves significantly contribute to the module's overall weight and their distribution on the outer walls makes the orientation control of the module challenging. The next major modification is a consolidation of skeleton walls and nozzles. The skeleton walls of TPODS module have specific mounting surfaces for the four nozzles, enabling precise and robust mounting. This also ensures that the nozzles are at the correct angle and distance from the center, ensuring the accurate simulation analysis. 

However, the pneumatic connection between control valves and nozzles using flexible tubes was a significant challenge during assembly. The placement of control valves inside the module is primarily driven by the size constraints. In contrast, the location of nozzles is driven by the their effectiveness to provide sufficient reaction forces. Consequently, mounting locations of control valves and nozzles can not be iterated for a suitable interconnection. This resulted in an interconnection path where the tube has to be bent in an `S' shape within a tightly constrained space. The initial attempt to achieve this failed due to the rigidness of the tube, resulting in an obstruction in airflow near bends due to the pinching of the tube. To overcome this, the tube is first inserted over an `S' shaped metal rod and then treated with a stream of hot air. Such treatment introduces some form of `memory' to the internal tube structure and, when cooled down, retains the `S' structure. Other notable modifications include an integrated circuit board with mountings for the on-board computer, Wi-Fi module and driver circuit for the solenoid valves. The power system has also been revamped and now includes a power on LED indicator and a switch to isolate power from the internal electronics completely. 

The improved design has also been replicated in the build process of a second TPODS module. In addition to that, Figure \ref{fig:TPODS_MKIII} shows a semi-rigid structure constructed from wooden sticks and aluminium 20-20 section. The mock-up of an external body resembles a truss structure with a central load. The mock-up also contains an adapter with magnets on both ends. The respective faces of the TPODS contain a matching pair of magnets such that the mock-up can be easily attached to TPODS modules. Powerful neodymium magnets bear the weight of the external body and provide a semi-rigid connection between two TPODS modules through the wooden sticks. It has been observed from preliminary experiments that this connection provides sufficient binding forces to elicit constraints of the motion of individual TPODS connected via the mock-up of an external body. Moreover, the disturbance caused by the pneumatic tether was identified as the primary factor governing the motion of the TPODS module. Since the disturbance is not entirely deterministic, it was decided to eliminate the pneumatic tether. As shown in Figure \ref{fig:TPODS_MOI}, PET accumulators are added. Although this limits the total data collection duration, a few of the adverse effects due to this have been mitigated by logging the pose at a higher rate than the simulation analysis. 

\subsection{Measurement of true inertial parameters}
TPODS module in its untethered configuration and both accumulators pressurized at their operating pressure, as shown in Figure \ref{fig:TPODS_MOI}, was placed on a weighing scale and the indicated weight is recorded. The accuracy of the scale is $\pm 2.5g$ for the selected operating range and the displayed weight of the module is $720 g$. Hence, it can be concluded that the true weight of the module is known with an accuracy of $0.7 \%$. The swigging test was conducted to measure the moment of inertia of the TPODS module\cite{MOI}. The module is mounted on two parallel strings, as shown in Figure \ref{fig:TPODS_MOI}. The oscillations in the axis perpendicular to the plane of motion are enforced and the resulting motion is carefully observed. The total time taken by the module to complete ten complete oscillation cycles are recorded. This activity is performed ten times to compensate for any errors during the initial perturbation force or in observing the duration. Finally, the time for each oscillation is computed by first averaging individual observations and then averaging ten collective observations. Once the oscillation time is computed, the moment of inertia is estimated using the following equation, where T is the time taken for one complete oscillation, D is the horizontal separation of the strings, h is the unsupported vertical length of the strings, m is the mass of the module and g is the acceleration due to gravity.

\begin{equation}
I_{zz} = \frac{mgD^2T^2}{16\pi^2h} 
\end{equation}

\begin{figure}[t!]
\begin{subfigure}[b]{0.38\textwidth}
    \includegraphics[width=\textwidth]{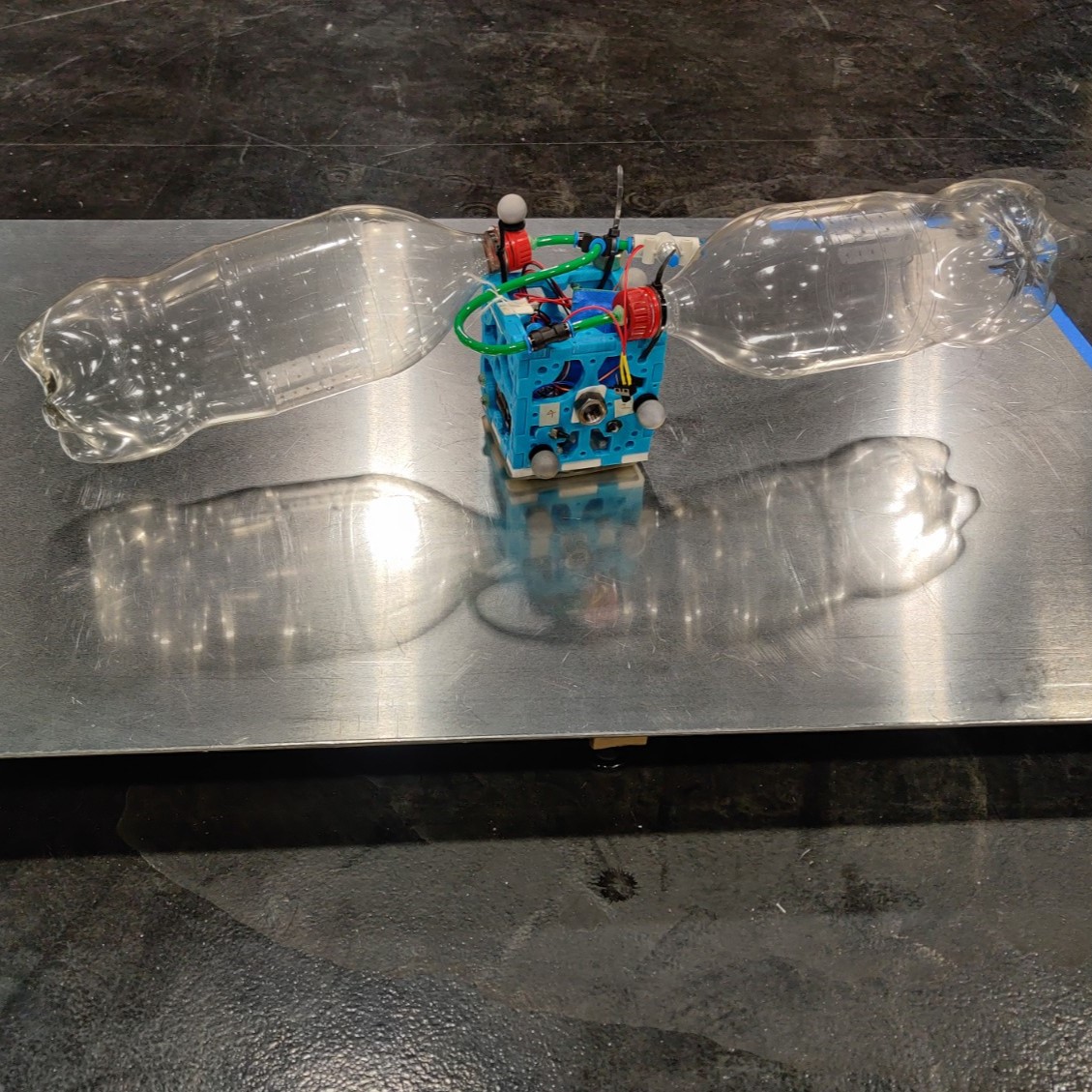}
\end{subfigure}
\hfill
\begin{subfigure}[b]{0.61\textwidth}
    \includegraphics[width=\textwidth]{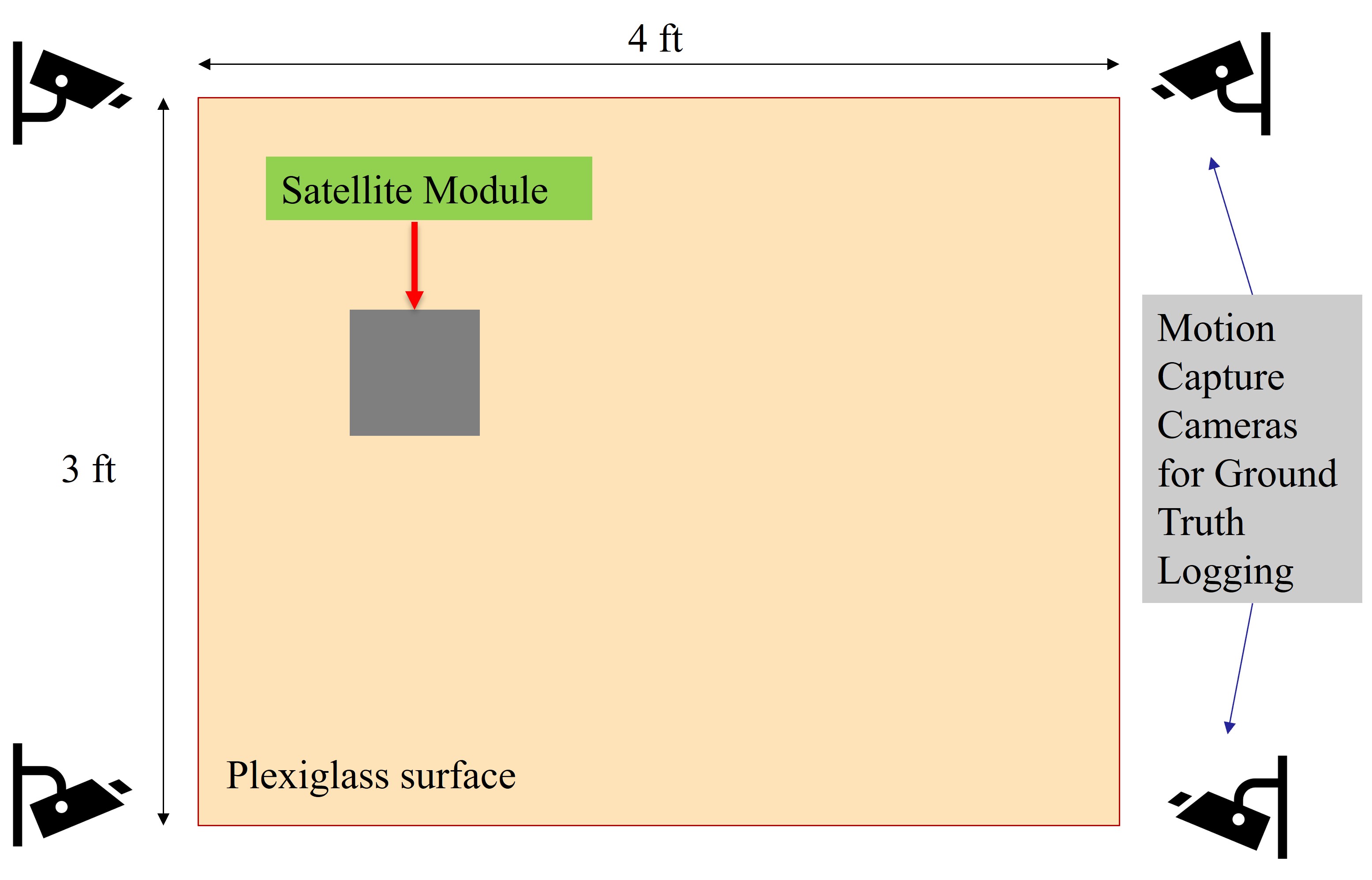}
\end{subfigure}
\caption{\label{fig:TPODS_marker}Pose measurement using VICON\textsuperscript{\tiny\textregistered} motion capture system}
\end{figure}

\subsection{Experimental Setup}
The intended goal of this experiment is to record the pose history of the TPODS module under the influence of predefined control inputs. It is important to note that although the final application of TPODS modules is detumbling of resident space objects, the mass and moment of inertia of the isolated TPODS module are being estimated first as a proof of concept. Hence, for the experiment, an isolated TPODS module is retrofitted with reflective markers such that they are not co-planar and spans the 3D volume. Figure \ref{fig:TPODS_marker} shows a TPODS module with reflective markers placed on a smooth surface within the range of multiple motion capture camera sensors. The surface is mounted on top of support structures with level adjustment mechanism. This is done to ensure that the surface on which the TPODS moves is completely level and does not introduce any bias in measurements. 

The module uses pressurized air stored inside the PET accumulators to move around. However, it is vital to ensure that during the experiment we maintain a reasonable pressure in the accumulators. As presented in earlier section, the thrusters are characterized at 60 PSI. Hence, operating the module around this pressure is essential to estimate the applied force accurately. Moreover, as concluded via sensitivity analysis, the nature of applied input significantly affects the effectiveness of the estimation algorithm. Consequently, a fourteen seconds input sequence that moves the module in an oscillatory motion in translation for the first seven seconds and in rotation for the next seven seconds is selected. The resulting motion of an isolated TPODS module is then measured using VICON\textsuperscript{\tiny\textregistered} motion capture system at a logging rate of 120Hz for further analysis.

\subsection{Analysis of experimental data}
\begin{figure}[h!]
\begin{subfigure}[b]{0.51\textwidth}
    \includegraphics[width=\textwidth]{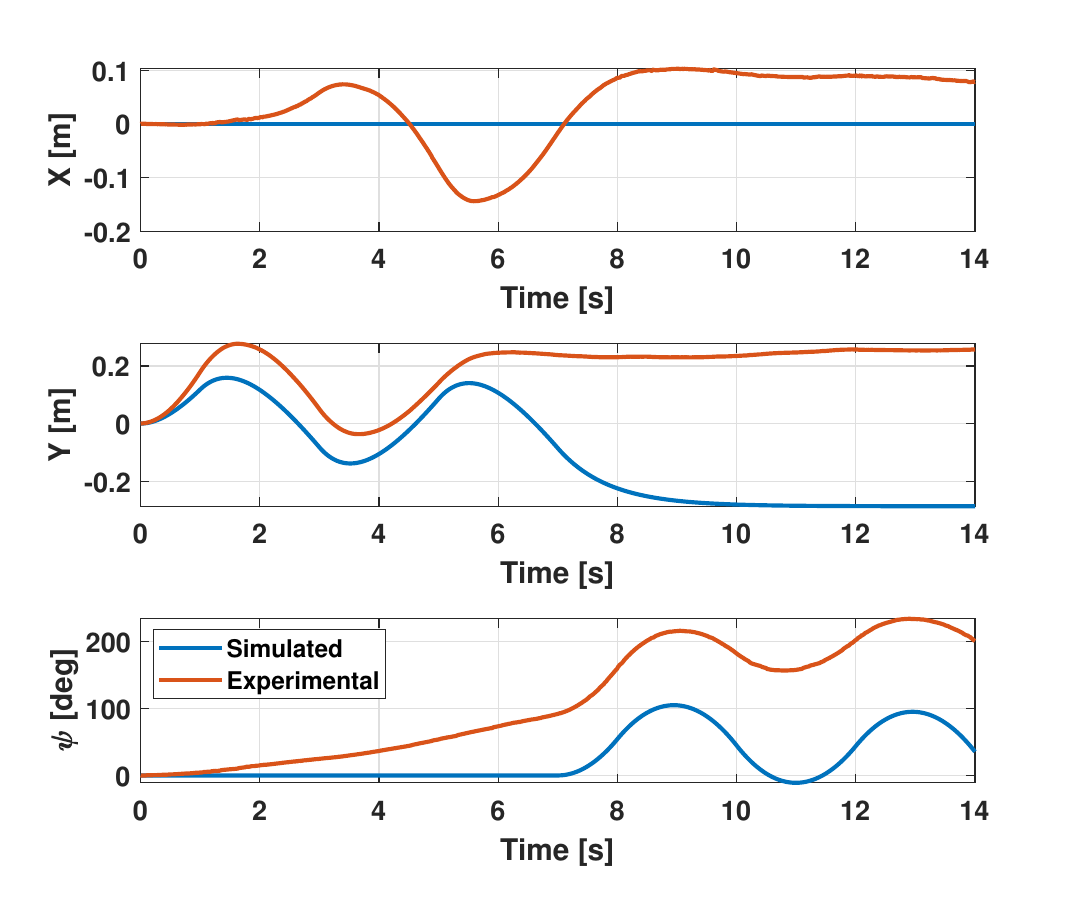}
    \caption{\label{fig:Ideal_actual}With friction model only}
\end{subfigure}
\hfill
\begin{subfigure}[b]{0.47\textwidth}
    \includegraphics[width=\textwidth]{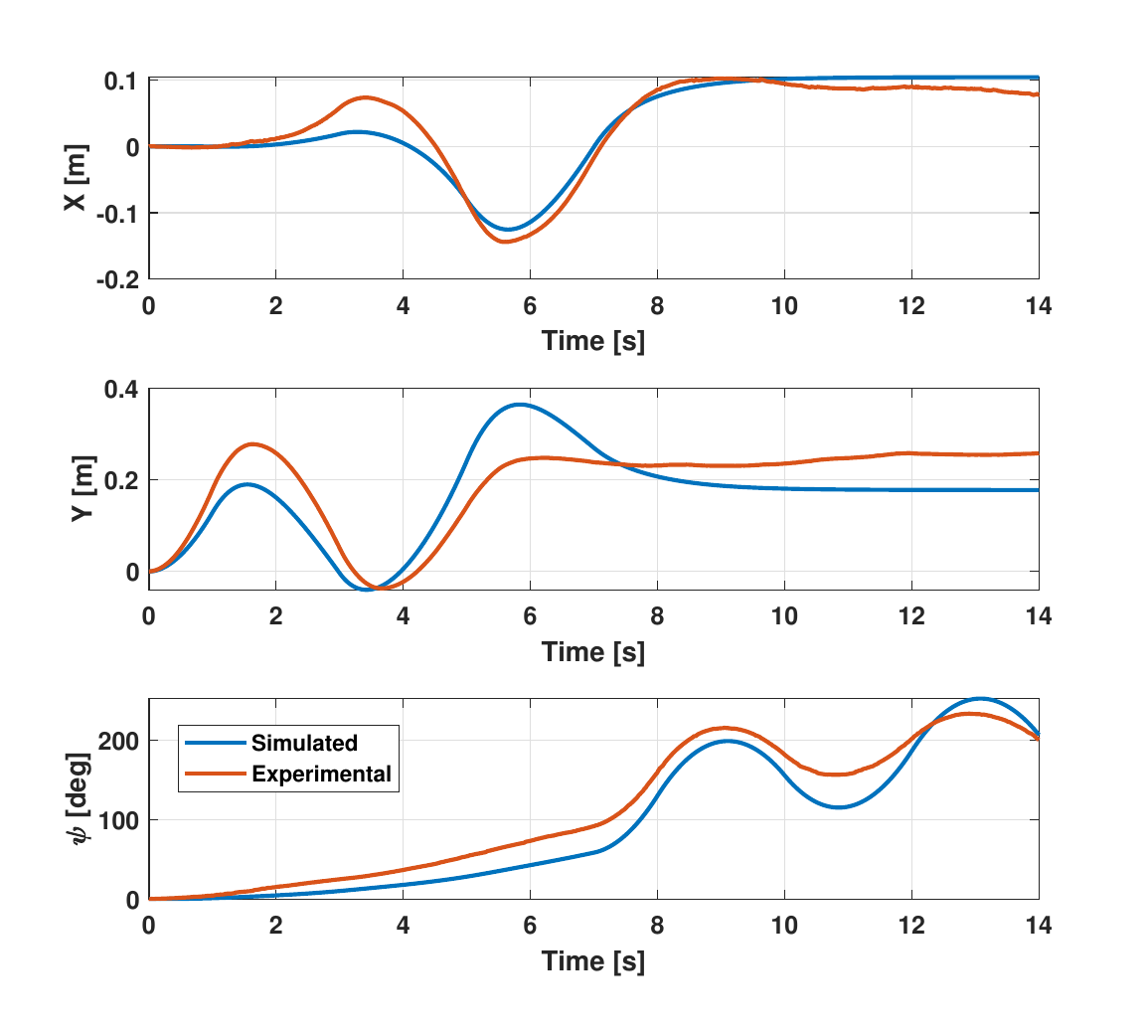}
    \caption{\label{fig:Ideal_actual_fric}With friction model and bias}
\end{subfigure}
\caption{Comparison between ideal and actual trajectory}
\end{figure}

The recorded pose data is further processed in MATLAB\textsuperscript{\tiny\textregistered} and analyzed for a matching with the simulation model. On comparing the recorded and simulated pose history under the same control inputs and inertial parameters, it was evident from the plots that there is a large deviation between simulated and actual trajectories. A particularly unusual behavior is a continuation of the oscillatory translation motion during the second half of the simulated trajectory. This is due to the accumulated momentum during the first half of the duration of motion. Since the simulation model assumes an ideal frictionless surface, there is no way to lose the accumulated momentum. However, we know that this is far from the truth. Hence, a frictional force is added to the simulation model. The friction is modeled as a velocity-dependent term directly opposing the translation and rotational motion of the module. The co-efficient of friction is adjusted to achieve a good matching between simulated and actual trajectory. Figure \ref{fig:Ideal_actual} shows the result of this modification. We can observe that the nature of both trajectories are in agreement and the translation stops during the second half, during which only rotational modes are excited. 

However, both trajectories are not quite matching and the probable reason for this is bias being introduced due to mechanical imperfections. The module has three point of contact to the surface, arranged at vertex of an equilateral triangle. In an ideal case, friction due to any translation motion should not induce rotational motion and vice versa. However due to mechanical imperfections, the module has a tendency to rotate in the positive direction while translating. In an ideal scenario, this should not be the case and the module should continue the motion under the influence of applied input and frictional forces only. To incorporate this behavior, a velocity dependent bias is added to the rotational dynamics. As shown in Figure \ref{fig:Ideal_actual_fric}, after adding a bias to rotational motion, simulated and experimental states are very close.  

\subsection{Extended Kalman Filter}
\begin{figure}[h!]
\begin{subfigure}[b]{\textwidth}
    \centering
    \includegraphics[width=\textwidth]{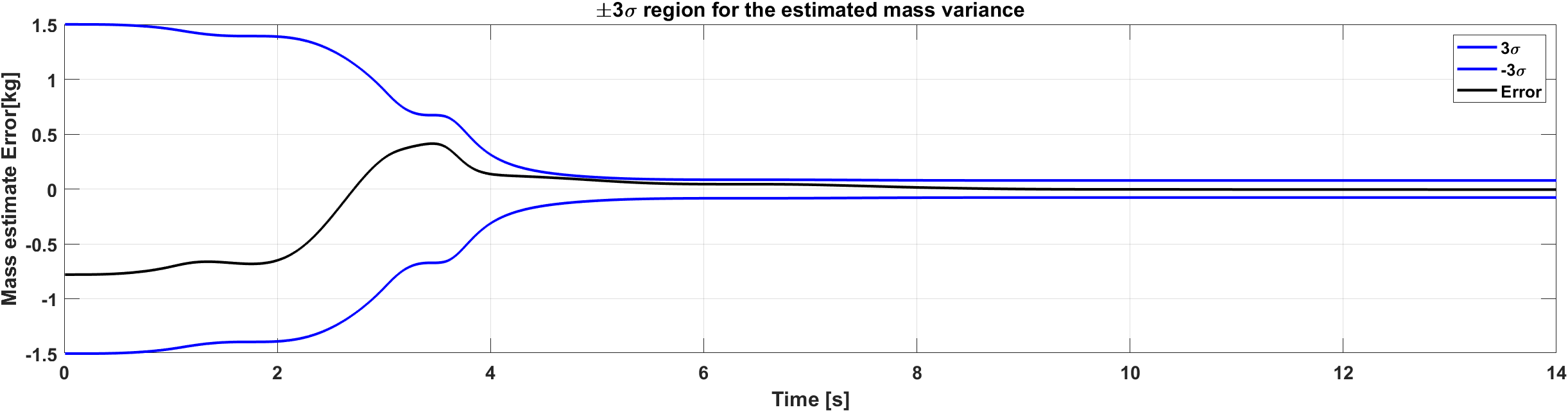}
\end{subfigure}
\begin{subfigure}[b]{\textwidth}
    \centering
    \includegraphics[width=\textwidth]{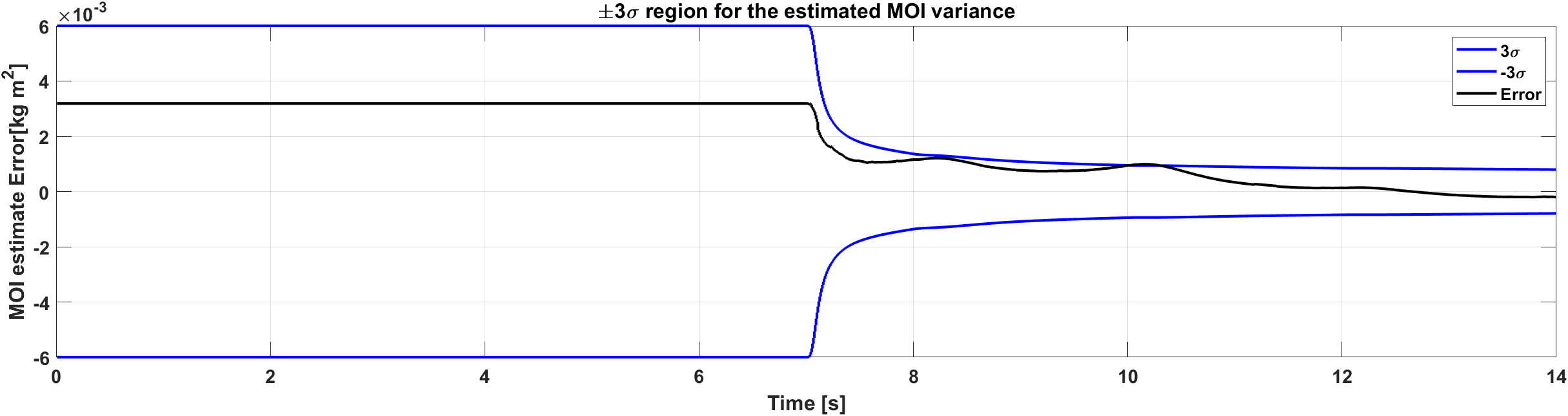}
\end{subfigure}
\caption{\label{fig:EKF_actual}Variance and error history of mass and MOI estimates using experimental data}
\end{figure}

The experimental pose data collected is fed to the Kalman filter-based estimation algorithm developed in the earlier section. Figure \ref{fig:EKF_actual} summarizes the variance regions and estimation error for mass and moments of inertia. It is evident from the figure that as the module starts moving, the variance of estimated mass begins to decrease, indicating that the algorithm is gaining confidence about the observed measurement based on the provided dynamics of the system. On the other hand, the variance in the moment of inertia estimate remains steady for the first half. This behavior is expected and can be attributed to the nature of applied input. Recall that the input sequence excites oscillatory modes in translation for the first half and rotation for the second. As soon as the module start rotating under the applied input, the variance of the moment of inertia estimate decreases and the estimate approaches the true value. Although the variance and error plots look visually correct, it is important to note that a significant amount of process noise has to be injected to ensure that the variance history stays inflated and encompasses the estimation error. This is due to the fact that we have a large number of highly accurate measurements but poorly known system dynamics. In addition to this, the coefficient of friction is also modified iteratively to produce desirable results. The Root Mean Square Error is 0.3050 kg for mass estimation and $2.32\times10^{-3}kgm^2$ for moment of inertia estimation. The increased RMSE for MOI estimation is attributed to the nature of applied input. 

\section{Conclusion}
A comprehensive model-based system design has been executed to realize the conceptual design into a working experimental setup. Ample analysis of the dynamical behavior of the TPODS module significantly aided the estimation process. The observability analysis unearthed an efficient input sequence, resulting in an accurate estimation of mass and moment of inertia using non-linear batch least squares technique. Some of the drawbacks of parameter estimation using an extended kalman filter algorithm have been eliminated by jointly leveraging the strengths of non-linear least squares and kalman filter algorithms. The initial design of TPODS module is revised to bring forth better dynamic stability and close matching with simulation behavior. True values of the mass and moment of inertia of the TPODS module are computed using established techniques. Finally, the actual pose history of TPODS module under the influence of a predefined input sequence is recorded, and data is utilized to assess the performance of the estimation algorithm. Although the outcome could be better, the concept of operation and technical feasibility has been shown with an embryonic experimental setup. This activity produced a clear set of capabilities of the estimation algorithm, propelling the eventual application of moving structures with partially known inertial properties in space.

\section{Acknowledgements}
This work is supported by the Air Force Office of Scientific Research (AFOSR), as a part of the SURI on OSAM project ``Breaking the Launch Once Use Once Paradigm" (Grant No: FA9550-22-1-0093). Course instructor Prof. Kyle DeMars and project mentor Prof. Manoranjan Majji are gratefully acknowledged for their watchful guidance, motivation, technical support, and discussions.

\bibliographystyle{AAS_publication}   
\bibliography{./bib/refsm-astro}   
\end{document}